\documentclass[a4paper]{cas-sc}

\usepackage[numbers,sort&compress]{natbib}
\usepackage{booktabs}
\usepackage{array}
\usepackage{placeins}
\usepackage{xcolor} 
\AtBeginDocument{%
  \hypersetup{
    colorlinks=true,
    linkcolor=blue,  
    citecolor=blue, 
    urlcolor=blue    
  }%
}
\def\tsc#1{\csdef{#1}{\textsc{\lowercase{#1}}\xspace}}
\tsc{WGM}
\tsc{QE}

\begin{document}
\let\WriteBookmarks\relax
\def\floatpagepagefraction{1}
\def\textpagefraction{.001}

\shorttitle{}    

\shortauthors{}  

\title [mode = title]{Conflict Avoidance in Pedestrian Merging in Controlled Experiments by Variance Indicator}  

\tnotemark[1] 

\tnotetext[1]{} 

%

\author[1]{Jiawei Zhang}



\ead{zhang-jiawei@g.ecc.u-tokyo.ac.jp}


\credit{Conceptualization, Formal analysis, Data curation, Funding acquisition, Methodology, Software, Visualization, Writing – original draft}

\affiliation[1]{organization={Department of Advanced Interdisciplinary Studies, School of Engineering, The University of Tokyo},
           addressline={4-6-1 Komaba, Meguro-ku}, 
           city={Japan},
          citysep={}, 
            postcode={153-8904}, 
            state={},
            country={Tokyo}}





\author[2]{Xiaolu Jia}
\affiliation[2]{organization={Beijing Key Laboratory of Traffic Engineering, Beijing University of Technology},
            addressline={No. 100 Pingleyuan, Chaoyang-district}, 
            city={Beijing},
          citysep={}, 
            postcode={100124}, 
            state={},
            country={China}}
\credit{Funding acquisition, Supervision, Writing-review}

\author[3]{Sakurako Tanida}
\credit{Funding acquisition, Supervision, Writing-review}
\ead{u-tanida@g.ecc.u-tokyo.ac.jp}
\author[3]{Claudio Feliciani}
\credit{Data curation, Funding acquisition, Investigation, Methodology, Supervision, Writing-review}
\affiliation[3]{organization={Department of Aeronautics and Astronautics, School of Engineering, The University of Tokyo},
            addressline={7-3-1 Hongo, Bunkyo-ku}, 
            city={Tokyo},
           citysep={}, 
            postcode={113-8656}, 
            state={},
           country={Japan}}

\author[3,1,5]{Daichi Yanagisawa}
\credit{Funding acquisition, Supervision, Writing – review and editing}

\author[3,1,4,5]{Katsuhiro Nishinari}
\credit{Conceptualization, Funding acquisition, Supervision}
\affiliation[4]{organization={Research Center for Advanced Science and Technology, School of Engineering, The University of Tokyo},
            addressline={4-6-1 Komaba, Meguro-ku}, 
            city={Tokyo},
          citysep={}, 
            postcode={153-8904}, 
            state={},
            country={Japan}}
            
\affiliation[5]{organization={Mobility Innovation Collaborative Research Organization, The University of Tokyo},
            addressline={}, 
            city={Tokyo},
          citysep={}, 
            state={},
            country={Japan}}          
            
\cortext[1]{Corresponding author}

\fntext[1]{0009-0004-4920-0131}


\begin{abstract}
Pedestrian congestion at corridor intersections often originates from localized fluctuations in motion rather than from a macroscopic collapse of flow. Understanding pedestrian instability at corridor intersections remains challenging because existing studies mainly rely on density, average speed, or flow-based measures and limited datasets, making it difficult to separate geometric turning effects from interaction induced fluctuations in merging flows. In particular, the mechanism underlying the turning angle dependence in T junctions has not been resolved. Here, we analyze more than 300 controlled experiments conducted in L corridors with turning only and T corridors with turning and merging. Using Voronoi-based speed variance $V_s$ and velocity variance $V_v$, we systematically compare geometric and interaction effects. $V_s$ effectively captures interaction driven instability, while $V_v$ reflects directional adjustments due to geometry. The comparison reveals distinct fluctuation mechanisms and identifies a critical transition near 90°, demonstrating the advantage of variance-based indicators for diagnosing pedestrian dynamics. 
\end{abstract}





\maketitle

\section{Introduction}
Traffic congestion, characterized by the accumulation and complex interactions of moving entities, represents a pervasive challenge across multiple scales and domains. Whether concerning satellites and space debris in orbit~\cite{Polli2023}, aircraft in flight corridors~\cite{Aditya2024}, maritime vessels in busy waterways~\cite{Kang2022}, or trains~\cite{Wang2025} and autonomous vehicles~\cite{Mushtaq2021} sharing limited tracks or roads, congestion arises universally, leading to safety risks, delays, and increased operational costs. To address these issues, extensive research has been conducted within each mode of transportation.

At the pedestrian scale, congestion presents distinct yet critical challenges, especially in dense urban settings, public venues, or emergency evacuations, where even minor disruptions can rapidly escalate into serious hazards. To mitigate the risks associated with pedestrian congestion, including stampedes, crushing incidents, or evacuation delays, substantial research has focused on understanding the underlying dynamics of pedestrian movement and interaction.

Crowd disasters frequently arise from complex, multi-directional pedestrian movements, particularly at locations where multiple pedestrian streams merge or intersect. Such merging behaviors, characterized by intersecting pedestrian flows, notably elevate the risk~\cite{Helbing2007,Kayvan2014} of congestion and consequent safety hazards.

However, in large intersections encountered in practice, such as urban plazas or open crossings, merging behavior is rarely governed by a single identifiable factor. Under these conditions, the simultaneous influence of corridor geometry, visual anticipation, social conventions, and heterogeneous walking intentions makes the notion of merging itself difficult to define precisely and operationally.

Previous studies have documented numerous incidents in which merging behaviors directly contributed to severe crowd disasters, underscoring the urgent need for a thorough understanding and improved management of these scenarios~\cite{Helbing2007,Kayvan2014}. Moreover, congestion’s occurrence and severity are heavily influenced by specific geometric constraints. Research indicates a marked decline in crowd density and flow efficiency at bottlenecks, intersections, and merging corridors, highlighting the significance of corridor geometry in shaping pedestrian dynamics~\cite{Shiwakoti2015,Shahhoseini2018}. In this paper, the geometric effect refers to the turning demand imposed by the junction layout, including the required change in walking direction and the available space to execute the turn, rather than to the mere existence of corridor walls.

The remainder of this paper is organized as follows. Section 2 discusses the relevant literature on corridor geometry behavior and indicators of pedestrian congestion and research gaps. Section 3 introduces the variance-based indicators used in this study. Section 4 describes the experimental setup and data acquisition. Section 5 presents the results and discussion, including a comparison between L and T corridor configurations, the spatial structure of merging dynamics in the T corridor, and the dependence on turning angle. Section 6 concludes the paper.

\section{Literature review}
Corridor geometry imposes constraints on pedestrian trajectories, and turning emerges as a typical adaptation to such constraints, often leading to changes in walking behavior.
Physiological studies~\cite{Courtine2003} have shown that when pedestrians negotiate a turn, their walking behavior is accompanied by systematic changes in body posture and inter-segmental coordination.
During curved walking, the head and trunk gradually reorient toward the turning direction, the body center of mass shifts inward, and foot placement is adjusted asymmetrically between the inner and outer legs.
These postural and kinematic adaptations increase movement complexity and are typically associated with a reduction in walking speed compared with straight path locomotion. 

Recent work by van der Vleuten \textit{et al.}~\cite{Vleuten2024} further analyzed stochastic fluctuations of diluted pedestrians walking along curved paths and showed that path curvature not only reduces mean walking speed but also systematically modulates longitudinal and transverse speed fluctuations, which a Langevin-type social-force model can capture.

Similarly, Ye \textit{et al.}~\cite{Ye2019} demonstrated that in turning scenarios, the highest pedestrian density typically appears at the inner corner, while the highest average speed occurs at the outer corner. This indicates that pedestrians prefer the shortest path when turning; however, this tendency often reduces their passing speed.

Moreover, the turning angle plays a crucial role in determining pedestrian dynamics. Experiments by Dias \textit{et al.}~\cite{Dias2014} revealed that the average walking speed decreases as the turning angle increases. Studies with different crowd sizes further suggest that the average speed of a group is generally lower than that of an individual, implying that inter-personal interactions contribute to speed reduction. In addition, Hannun \textit{et al.}~\cite{Hannun2022} reported that when a turning area is subdivided into smaller regions, speed reduction is most pronounced in the middle of the bend, especially under conditions of high desired walking speed and large turning angles. These findings were obtained in pure turning scenarios, where pedestrians negotiate a bend without any merging interactions. However, in the presence of merging, turning is no longer an isolated geometric process, and its impact on pedestrian dynamics may follow mechanisms that differ from those observed in pure turning scenarios.

Their results highlight that even in the absence of strong interactions, curved geometry alone already imprints a characteristic structure on microscopic variability in pedestrian trajectories.

However, in the presence of merging, turning is no longer an isolated geometric process.
Geometric constraints and interpersonal interactions become inseparably coupled, which complicates a direct definition of merging, but allows its effects to be inferred by comparing pure turning with turning accompanied by merging.

The study of pedestrian flow dynamics at merging junctions has drawn considerable attention due to its critical implications for safety and efficiency in urban environments. When two or more streams converge, pedestrians must continuously adjust their speed, direction, and spacing to negotiate priority and avoid collisions. Such adjustments inherently introduce fluctuations into the flow, and these fluctuations often become more pronounced than those caused by geometric features alone. Previous research has contributed significantly to understanding various aspects influencing congestion at junctions, although specific mechanisms underlying pedestrian dynamics are still not completely understood.

Complementing these macroscopic studies, Yu \textit{et al.}~\cite{Yu2026} conducted controlled experiments on pairwise collision-avoidance maneuvers in T-junctions with several merging layouts. By distinguishing head-on and perpendicular encounters and introducing the Minimum Predicted Distance (MPD) as a dynamic indicator of avoidance effectiveness, they showed that encounter type, passing order, and desired speed systematically shape individual avoidance strategies during merging.

Simulation studies that model collision forces arising from different insertion angles have shown that larger insertion angles intensify these interactions and reduce walking speeds~\cite{Wong2010}. Additional simulations that varied corridor curvature without inserting pedestrians revealed that straighter corridors tend to support more stable self-organized lanes and higher speeds, whereas the complexity introduced by merging disrupts this stability~\cite{Sharifi2020}. Moreover, although turning angles influence movement in one-way streams, this effect largely disappears in bidirectional flow, indicating that the interference generated by opposing pedestrians becomes dominant over geometric curvature.

Existing studies clearly establish several critical points. Zhang \textit{et al.}~\cite{Zhang2012} empirically examined pedestrian streams at T-junctions and found that congestion is more severe before merging occurs than after merging. Importantly, in scenarios involving only turning without merging, no notable difference in pedestrian flow was observed before and after the corner. This suggests that the geometric act of turning alone does not substantially alter macroscopic flow characteristics, and that the severe congestion observed in T-junction experiments must therefore arise primarily from merging interactions rather than from turning itself. Aghabayk \textit{et al.}~\cite{Kayvan2015} extended this line of inquiry by investigating how merging angles affect pedestrian behavior. They concluded that pedestrian speeds and flow rates exhibit a nonlinear relationship with merging angles. Notably, asymmetric configurations, particularly at a 90-degree angle, demonstrated significantly worse congestion than more symmetric angles like 60 and 120 degrees. Lian \textit{et al.}~\cite{Lian2017} further explored the influence of feeder lane width on pedestrian merging behavior, concluding that wider feeder lanes intensify bottleneck phenomena between main and merging streams. Yu \textit{et al.}~\cite{Yu2018} further examined the role of merging layout in pedestrian dynamics and showed that different geometric arrangements can significantly influence merging performance, with larger merging angles generally leading to lower efficiency. Unlike earlier studies, they introduced novel indicators, including lane entropy to quantify the disorder of lane formation and a harmony index to characterize positional disharmony among pedestrians.

Taken together, these studies indicate that turning and merging involve fundamentally different mechanisms. Turning primarily imposes kinematic constraints on trajectories and visibility, while merging introduces direct competition for space and priority, leading to strong interpersonal interference. At the same time, results on the effect of turning angle in merging corridors are not fully consistent: as the angle increases, walking speed is generally expected to decrease, yet some experiments report higher flow at  120° than 90°, apparently contradicting the monotonic trend suggested by pure turning experiments. While previous studies have extensively examined geometric factors in merging scenarios, such as merging angles and corridor widths, the dynamic processes underlying congestion in these configurations remain insufficiently characterized. In particular, although variations in speed and density have been widely reported, existing analyses often rely on macroscopic or averaged measures that provide limited insight into how pedestrian interactions dynamically evolve during merging. As a result, a key challenge in merging studies lies not only in understanding geometric effects, but also in developing appropriate quantitative indicators that can capture the localized and interaction-driven nature of congestion. 

Accordingly, a second major line of research focuses on how pedestrian performance and congestion in merging scenarios can be effectively quantified. Predominantly, existing indicators such as Kernel Density Estimation (KDE) primarily measure static pedestrian density distributions, lacking insights into pedestrian velocity dynamics. The Level of Service (LOS) metric categorizes congestion levels from A to F~\cite{fruin1971}, yet does not reflect pedestrians’ dynamic behaviors, failing to distinguish static crowds from dynamically interacting pedestrians. 
Density and velocity fields are often constructed by projecting discrete pedestrian trajectories onto fixed spatial grids. Such grid-binned representations can introduce numerical artifacts, including artificial jumps when pedestrians cross cell boundaries, leading to discontinuities in the estimated fields in both time and space. Kernel-based methods have therefore been widely adopted to smooth discrete data and obtain continuous density and velocity fields.
However, as pointed out by Duives \textit{et al.}~\cite{Duives2015}, elocity does not necessarily exhibit a monotonic relationship with density, indicating that even smoothed macroscopic fields may fail to capture the underlying dynamics of pedestrian flow. 
These issues complicate the interpretation of fundamental diagrams and reduce the reliability of density–velocity relationships for diagnosing local congestion.

Dynamic measures, such as Congestion Level~\cite{Feliciani2018} and Congestion Number~\cite{Zanlungo2023}, utilize vorticity and its spatial derivatives to characterize pedestrian disorder. These indicators capture swirling, blocking, and recirculation patterns in the velocity field and have been successfully applied to identify intrinsically risky regions in crowds. However, they focus primarily on group-level dynamics without adequately capturing individual interactions, and they are most effective when movement directions are highly dispersed. Entropy measures provide another approach. Speed or velocity based entropy~\cite{Huang2015,Xie2022,Wang2018,Zeng2021}, typically partitions the joint distribution of speed magnitude and walking direction into discrete bins and computes Shannon entropy as a measure of disorder. In this framework, a speed entropy component reflects the dispersion of speed magnitudes across bins, while a direction entropy component quantifies how widely headings are spread. Their product or combination yields a velocity entropy that increases when pedestrians move with heterogeneous speeds and directions. In parallel, Spatial Aggregation Entropy~\cite{Guo2020}. Acceleration Quantile (AQ)~\cite{Wang2018} indicators incorporate dynamic properties at the individual level but are disproportionately sensitive to outliers, thus potentially introducing significant inaccuracies. Additionally, aggregation or entropy-based measures often highlight broad regions of elevated disorder without clarifying which specific micro mechanisms. Speed adjustments versus directional changes contribute most to observed congestion.

While recent advances using machine learning methods, such as Bidirectional Long-short-term-memory (Bi-LSTM) architectures~\cite{Basalamah2023}, offer promising directions, these approaches require extensive data and tend to obscure the underlying physical mechanisms driving pedestrian dynamics. Consequently, their outputs can be difficult to interpret and to relate directly to geometric design variables or interaction rules. Thus, despite the variety of available indicators, within the field of pedestrian traffic flow modelling there does not seem to be consensus on the question which of these measures performs the best, either in terms of robustness or in their ability to capture the early, localized fluctuations that precede macroscopic congestion.

Consequently, there remains an essential gap in understanding how pedestrian interactions explicitly contribute to congestion and mobility bottlenecks in merging scenarios and how these interaction effects compare with the geometric effects observed in pure turning scenarios. While earlier studies have examined geometric configurations such as merging angles and corridor widths, substantial gaps remain in understanding the dynamic mechanisms driving congestion, especially concerning localized pedestrian interactions and related speed fluctuations.
An effective indicator for merging dynamics should not only be sensitive to local interactions but also remain interpretable and comparable across different geometric configurations.

To address these unresolved challenges, this study proposes a physically interpretable indicator tailored for corridor environments and applies it systematically to both turning and merging scenarios. 
The new measure is based on local speed variance within Voronoi-defined neighborhoods and aims to directly capture pedestrian movement states and elucidate interaction mechanisms, thereby filling a critical gap in existing pedestrian dynamics research. By conducting paired experiments in an L corridor (turning only) and a T corridor (turning plus merging) with comparable geometric parameters and crowd sizes, we obtain a consistent basis for separating the effects of geometry from those of merging interactions.

Given these dynamics, this study specifically aims to capture local heterogeneity by focusing on variations in pedestrian velocities around the turning and merging areas. Actively recognizing and utilizing pedestrians’ subjective awareness in managing their speed enables effective conflict avoidance and promotes self-organized, orderly movement patterns. In addition, this understanding guides optimizing corridor design to reduce environmental obstructions and mitigate unnecessary interaction-induced fluctuations. The ultimate aim is to identify the underlying dynamical mechanisms of merging under controlled geometric conditions, with the prospect that these mechanisms can help explain more complex merging situations in which geometric constraints are less explicit.This understanding can support identifying high risk merging conditions and improving safety management.

\section{Variance indicators}\label{sec:pf}
In this study, we use Voronoi diagrams as a fundamental approach to categorize pedestrian interactions in merging scenarios. A Voronoi diagram divides space into distinct regions, or Voronoi cells, each corresponding to an individual pedestrian. Each cell comprises all points closer to its associated pedestrian than to any other, with boundaries defined by the perpendicular bisectors of lines connecting neighboring pedestrians. The area of a Voronoi cell thus represents the available personal space for movement~\cite{Steffen2010,Jia2022}. Two pedestrians are considered neighbors if their Voronoi cells share the same boundary, and subsequent computations are based on these neighboring relationships.

\begin{figure}
\centering
\includegraphics[width=0.4\textwidth]{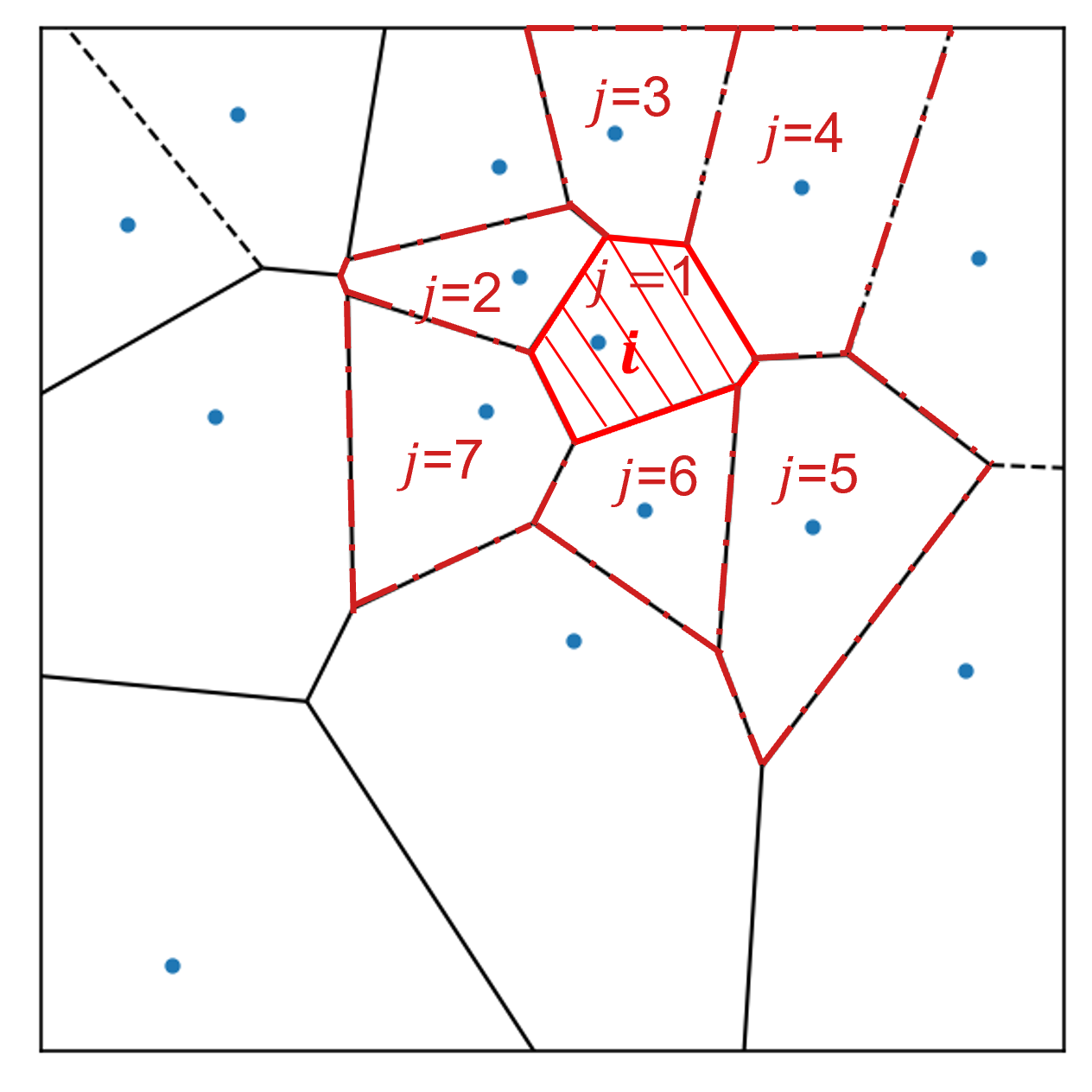}
\caption{\textbf{Illustration of the Voronoi-based definition of local regions and neighbors for pedestrian $i$.} Blue dots represent pedestrian centroid positions (generator points) extracted from the trajectory data. The plane is partitioned into Voronoi cells (black polygons), where each cell contains all points closer to its associated pedestrian than to any other. The highlighted red cell denotes the Voronoi region $V_i$ of the target pedestrian $i$. Surrounding red-outlined polygons indicate the Voronoi neighbors $\{\, j \in \mathcal{N}_i \,\}$, defined as pedestrians whose cells share a common boundary with $V_i$. These neighbors ($j=1, 2,3,4,5,6,7$ in the figure) form the local interaction set used in the computation of speed variance $V_s$ and velocity variance $V_v$. The shared edges represent the perpendicular bisectors of pedestrian pairs, forming the physical competition boundaries for available space.
\label{voronoi}}
\end{figure}

Taking individual $i$ as an example (FIG.~\ref{voronoi}), the corresponding Voronoi cell is referred to as the “cell of individual $i$.” The cells of individuals $j$ =1, 2, 3, 4, 5, 6, and 7 (including themselves) share boundaries with this cell and are therefore considered as the Voronoi neighbors of individual $i$.
To compute the speed variance $V_s$ for the cell of individual 1, a set of individuals is formed, consisting of the target individual $i$ and all its Voronoi neighbors ($j$ = 1, 2, 3, 4, 5, 6, 7). The speed values of this group are used to calculate the variance, which is then assigned to the Voronoi cell of individual $i$.

This process is repeated for each Voronoi cell in the scene, assigning a speed variance $V_s$ value to every cell, which serves as the basis for subsequent analysis.

Initially, it is essential to distinguish genuine fluctuations in pedestrian movement caused by merging interactions from simple deceleration effects, such as speed reductions that occur solely due to turning at intersections. If pedestrian velocities, which are vector quantities, are used directly, directional changes at corridor turns may lead to substantial fluctuations. These arise from sign changes in velocity components (positive to negative or vice versa), even when the actual changes in speed are minimal. To prevent such a misinterpretation, we introduce a variance measure based solely on speed magnitude, which is a scalar quantity, called speed variance $V_s$:
\begin{eqnarray}
\mathrm{Var}_s^{(j)} &=& \frac{1}{|\mathbf{I}_j |\bar{v}} \sum_{i \in \mathbf{I}_j} (v_i - \bar{v})^2 ,
\end{eqnarray}
\begin{eqnarray}
V_s=\mathrm{Var}_s^{(j)}
\end{eqnarray}
where $\mathbf{I}_j$ is  is the SET of Volonoi neighbor of individual $j$ and $\mathbf{I}_j$ denotes the number of Voronoi neighbors.

For each pedestrian $i \in \mathbf{I}_j$, define the instantaneous velocity vector as $\vec{v}_i = (v_{ix}, v_{iy})$. 
Here, $v_i$ represents the speed, $v_i$ = $\sqrt{v_{ix}^2 + v_{iy}^2}$, which is the scalar magnitude of velocity, of individual $i$ computed by averaging instantaneous velocities over 15 consecutive frames, with each frame lasting approximately 0.03 seconds. The choice of a 15-frame interval is based on two considerations: shorter intervals result in minimal pedestrian displacement~\cite{Jia2019}, while longer intervals may obscure subtle fluctuations in speed. These minor fluctuations may reflect pedestrian hesitation or small behavioral adjustments and are important to capture accurately. Since pedestrian trajectories are tracked using head positions rather than foot placements, measurement errors caused by natural foot oscillations during walking are effectively avoided.
The term $\bar{v}$ denotes the average speed of the target individual and all pedestrians whose Voronoi cells share a boundary with the target cell. The number of individuals included in each calculation, denoted by $N$,  varies locally depending on the immediate spatial configuration, and includes the target individual itself. Once calculated, the speed variance $V_s$ is assigned to the corresponding Voronoi cell.

To explicitly quantify directional heterogeneity, we define the velocity variance $V_v$ as:

\begin{equation}
\mathrm{Var}_v^{(j)} = \frac{1}{|\mathbf{I}_j |\bar{\vec{v}} } \sum_{i \in \mathbf{I}_j} \left\| \vec{v}_i - \bar{\vec{v}} \right\|^2 ,
\label{eq:velocity_variance}
\end{equation}

\begin{eqnarray}
V_v=\mathrm{Var}_v^{(j)}
\end{eqnarray}

In Equation~\ref{eq:velocity_variance}, $\vec{v}_i$ represents the velocity vector of pedestrian $i$, and $\bar{\vec{v}}$ is the average velocity vector of the target individual and its Voronoi neighbors.  Therefore, $V_v$ captures the directional dispersion in pedestrian motion by accounting for both speed and directional variations at a local level.

To distinguish directional fluctuations from speed-related noise, we compute an angular variance measure that is invariant to speed magnitude. This allows us to assess whether the observed velocity variance $V_v$ is primarily driven by directional dispersion. So compute the turning angle:

\begin{equation}
\phi_i = \tan^{-1}\left( \frac{v_{iy}}{v_{ix}} \right), \quad \phi_i \in [-\pi, \pi]
\label{eq:angle_def}
\end{equation}

Because heading angles lie on the circle, Euclidean variance is inappropriate. Instead, we use the circular mean and mean resultant length to compute angular variance. The circular mean angle is defined as:

\begin{equation}
\bar{\phi}_j = \tan^{-1}\left( \frac{ \sum_{i \in \mathbf{I}_j} \sin\phi_i }{ \sum_{i \in \mathbf{I}_j} \cos\phi_i } \right)
\label{eq:circular_mean}
\end{equation}

The mean resultant length is:

\begin{equation}
R_j = \left| \frac{1}{|\mathbf{I}_j|} \sum_{i \in \mathbf{I}_j} e^{i\phi_i} \right|, \quad R_j \in [0,1]
\label{eq:mean_resultant_length}
\end{equation}

Finally, the circular variance is computed as:

\begin{equation}
\mathrm{Var}_\phi^{(j)} = 1 - R_j
\label{eq:angular_variance}
\end{equation}

\begin{eqnarray}
V_\phi=\mathrm{Var}_\phi^{(j)}
\end{eqnarray}

Here, $\mathrm{Var}_\phi^{(j)} = 0$ indicates perfect alignment (all pedestrians walking in the same direction), while $\mathrm{Var}_\phi^{(j)} \to 1$ implies maximum dispersion (uniformly distributed headings). This measure isolates directional heterogeneity and serves as a benchmark for evaluating the contribution of heading variation to overall velocity variance.

When comparing experimental data across different conditions, such as varying pedestrian numbers, turning angles, or merging configurations, the absolute magnitude of speed variance $V_s$ may be influenced by differences in the mean walking speed. To avoid confounding effects arising from variations in the speed scale and to ensure comparability among conditions, we employ a normalized form of the speed variance.
Specifically, $V_s$ is normalized by the square of the local mean speed $\bar{v}$, yielding the normalized speed variance $\mathrm{Var}(nor)_s^{(j)}$ as defined in Eq.~\ref{eq:normalize}. This normalization allows fluctuations in walking speed to be evaluated on a relative scale, facilitating a consistent comparison of dynamical variability across different experimental scenarios.

\begin{eqnarray}
\mathrm{Var}(nor)_s^{(j)} &=& \mathrm{Var}_s^{(j)}/\bar{v}^2
\label{eq:normalize}
\end{eqnarray}

\section{Experiment}\label{sec:pf}
In 2016 (June and October), a series of controlled experiments was conducted  at the University of Tokyo (Tokyo, Japan).

Specifically, two experimental scenarios were examined: pedestrian turning in an L-shaped corridor and pedestrian merging in a T-shaped corridor. The dataset comprises a total of 318 experimental trials, including 196 trials in the L-shaped corridor and 122 trials in the T-shaped geometry, providing a statistically rich basis for the subsequent analysis.

It should be noted that not all experimental trials are used in the subsequent analysis. The quantitative analysis focuses on experiments with 12, 24, and 40 pedestrians. For the L-shaped corridor, only trials in which all pedestrians turn left are considered.

Across both scenarios, the corridors were enclosed by cardboard acting as walls. A uniform corridor width of 2 m was used in all experiments, and the height of the (cardboard) walls was about 2.2 m, meaning that people were not able to see the exterior of the corridor from the inside. A starting area was created 8 m before the elbow of the corridor, where participants would gather before the start of each trial. The starting area itself had a length of 8 m, and people were randomly and uniformly distributed inside it before each trial. The stretch between the elbow and the waiting area was intended to let participants accelerate and reach a stable speed before reaching the turning or merging area. Participants were instructed to continue walking without stopping after exiting the corner (and the experimental area) through an additional 4 m segment delimited with walls initially and partitions later to ensure a fluid and uninterrupted movement within the measurement zone.

The primary aim of this experiment was to investigate the impact of varying merging angles on pedestrian merging behavior. Consequently, one leg of the corridor was bended at different angles ($\theta$ = 30°, 60°, 90°, 120°, and 150°) and pedestrian density in the starting area was varied.

In both sessions, at least 40 participants were hired for the experiments. All participants provided written informed consent before participating, and the experimental procedure was approved by The University of Tokyo before data collection.

A camera was mounted directly above the intersection of the two corridors at approximately 5.5~m above ground, recording video at a frame rate of 29.97 frames per second (FPS) and a resolution of 1920 $\times$ 1080 pixels. Calibration was performed using selected reference points within the experimental area to ensure consistency between the digital coordinate system of the camera and the actual physical space. Tracking was performed using colored hats as markers, and different caps color were used to distinguish between participants walking in the straight section of the corridor and those turning. The tracking software \textit{PeTrack}~\cite{petrack} was used to identify hat colors and extract pedestrian trajectories from the recorded video data. Pixel-to-physical coordinate conversion was based on the average height of participants.

The trial concluded when the last participant exited the corridor.
All participants wore black T-shirts with green and blue tape affixed to their left and right shoulders, respectively, to facilitate shoulder identification.

\begin{figure}[h]
\includegraphics[width=1\textwidth]{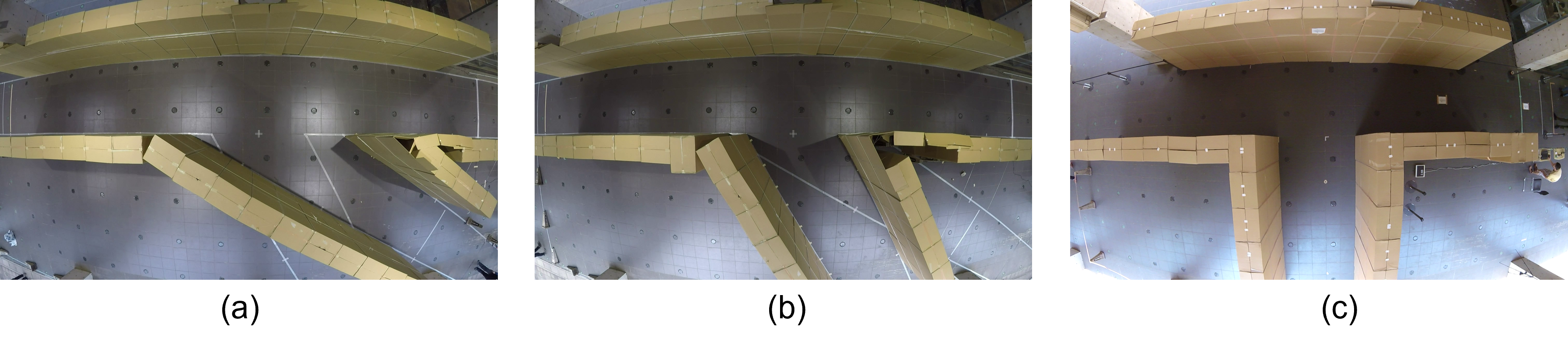}
\caption{\textbf{Empty corridor layouts used for constructing the turning and merging experimental geometries.}(a) Corridor configuration used to create the 30° and 150° turning scenarios. Pedestrians entering from the lower branch experience a 30° left turn, while reversing the direction yields a 150° right turn.
(b) Corridor configuration for generating 60° and 120° turning angles. A 60° left turn is formed for pedestrians approaching from the lower branch, and the same geometry produces a 120° right turn when traversed in the opposite direction.
(c)The corridor configuration corresponding to the 90° turning case.\label{empty corrdiors}}
\end{figure}

Pictures showing the empty corridors used in the experiments are presented in  FIG.~\ref{empty corrdiors}. In the following sections, more specific details will follow on the turning and merging experiments separately.

Experimental data containing trajectories and a detailed description of the geometrical layout are openly available for independent investigations~\cite{Feliciani2016}.

\subsection{Experiment procedure of L corridor}
\begin{figure}[h]
\centering
\includegraphics[width=0.3\textwidth]{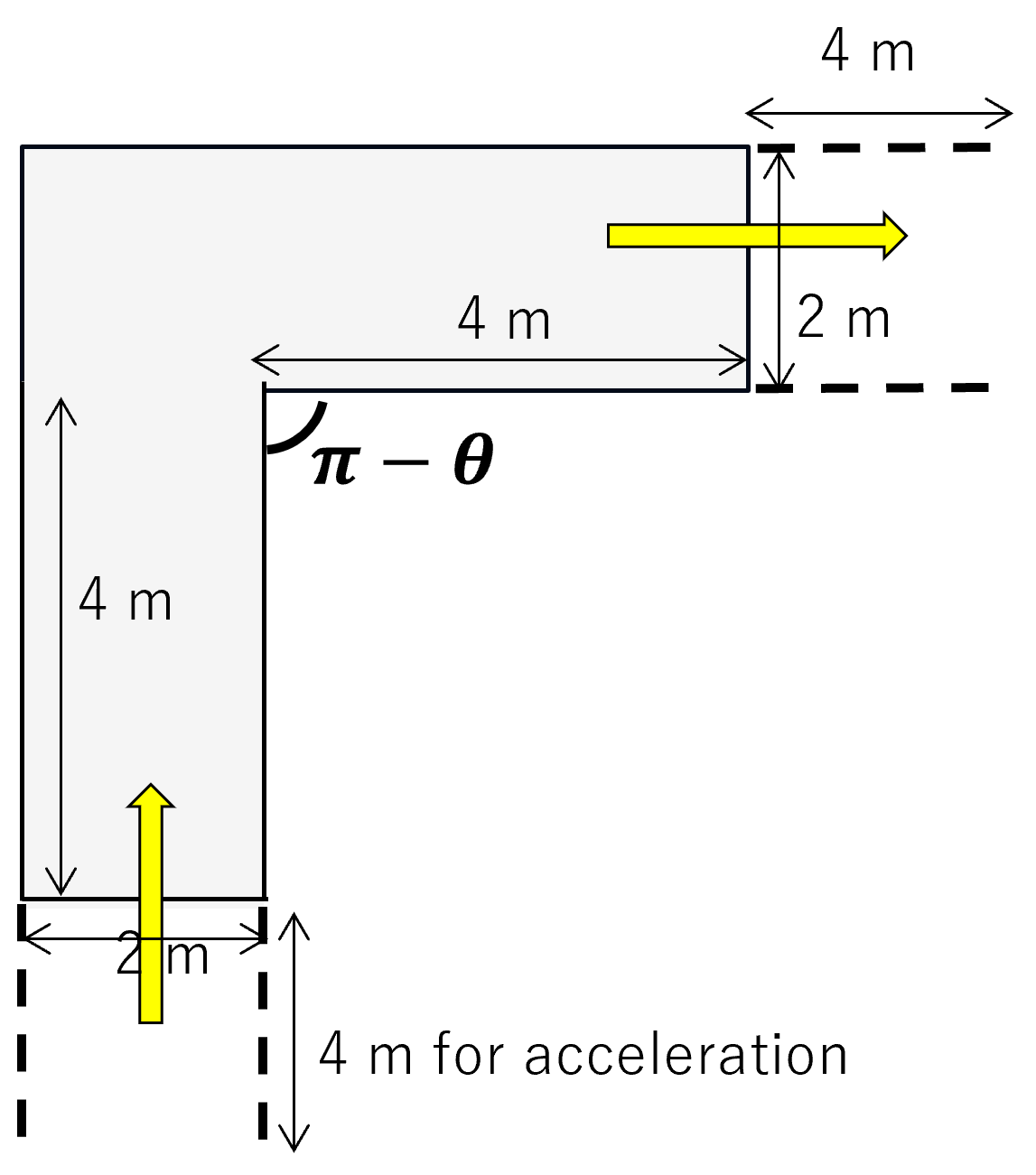}
\caption{\textbf{Experimental layout of the L corridor (turning-only scenario).} The corridor consists of two straight segments of equal width (2 m) connected by a turning angle, forming an L-shaped geometry. An 8 m start area and a 4 m acceleration zone are placed upstream to allow pedestrians to reach a stable walking speed before entering the measurement region. After negotiating the turn, pedestrians walk through an additional 4 m downstream segment to avoid premature deceleration.\label{setupL}}
\end{figure}

The layout of L corridor (FIG.~\ref{setupL}) consists of two straight segments, each 4 m long and 2 m wide. Between the waiting area and the walled section of the corridor, there is an additional 4 m acceleration zone to allow participants to reach a steady walking speed before entering the corner.

In L corridor, the set of pedestrian densities differed across angles (Table~\ref{tab:L_trial_config}). For 30° and 60°, four density levels were tested with 4, 12, 24, and 40 participants. For the 90° condition, eight density levels were examined, involving 1, 4, 8, 12, 16, 24, 32, and 40 participants. For 120° and 150°, seven density levels were tested with 4, 8, 12, 16, 24, 32, and 40 participants.


Combining these variables resulted in 30 experimental conditions. To ensure the reliability of the measurements and reduce random variability, each condition was repeated at least four times. Conditions with a small number of people were repeated more times, since a higher variability is expected in the results, and the higher sample size can help in gaining valid statistical results. The resulting dataset comprised 196 trials.
Table~\ref{tab:L_trial_config} summarizes the experimental configurations and trial parameters for the L corridor scenario, including the turning angle, the number of pedestrians (\#Pedestrian), design density in the starting area ($\rho$), number of repetitions (Rep), and turning direction.

\begin{table}[htbt]
  \centering
  \caption{
  Summary of all L corridor experimental configurations and trial parameters.}
  \label{tab:L_trial_config}
  \begin{tabular}{lccccc}
    \hline
    \textbf{Angle} & \textbf{Ped} & $\boldsymbol{\rho}$ & \textbf{Rep} & \textbf{Turning Direction} \\
    \hline
    30$^\circ$ & 4  & 0.25 & 7 & Left \\
    30$^\circ$ & 12 & 0.75 & 5 & Left \\
    30$^\circ$ & 24 & 1.5  & 5 & Left \\
    30$^\circ$ & 40 & 2.5  & 4 & Left \\
    60$^\circ$ & 4  & 0.25 & 7 & Left \\
    60$^\circ$ & 12 & 0.75 & 7 & Left \\
    60$^\circ$ & 24 & 1.5  & 5 & Left \\
    60$^\circ$ & 40 & 2.5  & 4 & Left \\
    90$^\circ$ & 1  & 0.0625  & 5 & Left \\
    90$^\circ$ & 4  & 0.25 & 6 & Left \\
    90$^\circ$ & 8  & 0.5  & 5 & Left \\
    90$^\circ$ & 12 & 0.75 & 5 & Left \\
    90$^\circ$ & 16 & 1.0  & 6 & Left \\
    90$^\circ$ & 24 & 1.5  & 6 & Left \\
    90$^\circ$ & 32 & 2.0  & 5 & Left \\
    90$^\circ$ & 40 & 2.5  & 6 & Left \\
    90$^\circ$ & 12 & 0.75 & 5 & Right \\
    90$^\circ$ & 24 & 1.5  & 4 & Right \\
    90$^\circ$ & 40 & 2.5  & 4 & Right \\
    120$^\circ$ & 4  & 0.25 & 12 & Left \\
    120$^\circ$ & 8  & 0.5  & 7 & Left \\
    120$^\circ$ & 12 & 0.75 & 6 & Left \\
    120$^\circ$ & 16 & 1.0  & 6  & Left \\
    120$^\circ$ & 24 & 1.5  & 5  & Left \\
    120$^\circ$ & 32 & 2.0  & 5 & Left \\
    120$^\circ$ & 40 & 2.5  & 5 & Left \\
    150$^\circ$ & 4  & 0.25 & 6 & Left \\
    150$^\circ$ & 8  & 0.5  & 7 & Left \\
    150$^\circ$ & 12 & 0.75 & 7 & Left \\
    150$^\circ$ & 16 & 1.0  & 7 & Left \\
    150$^\circ$ & 24 & 1.5  & 6 & Left \\
    150$^\circ$ & 32 & 2.0  & 10 & Left \\
    150$^\circ$ & 40 & 2.5  & 6 & Left \\ 
    \hline
  \end{tabular}
\end{table}
\FloatBarrier

\subsection{Experiment layout of the T corridor (turning with merging)}
\begin{figure}[h]
\centering
\includegraphics[width=0.5\textwidth]{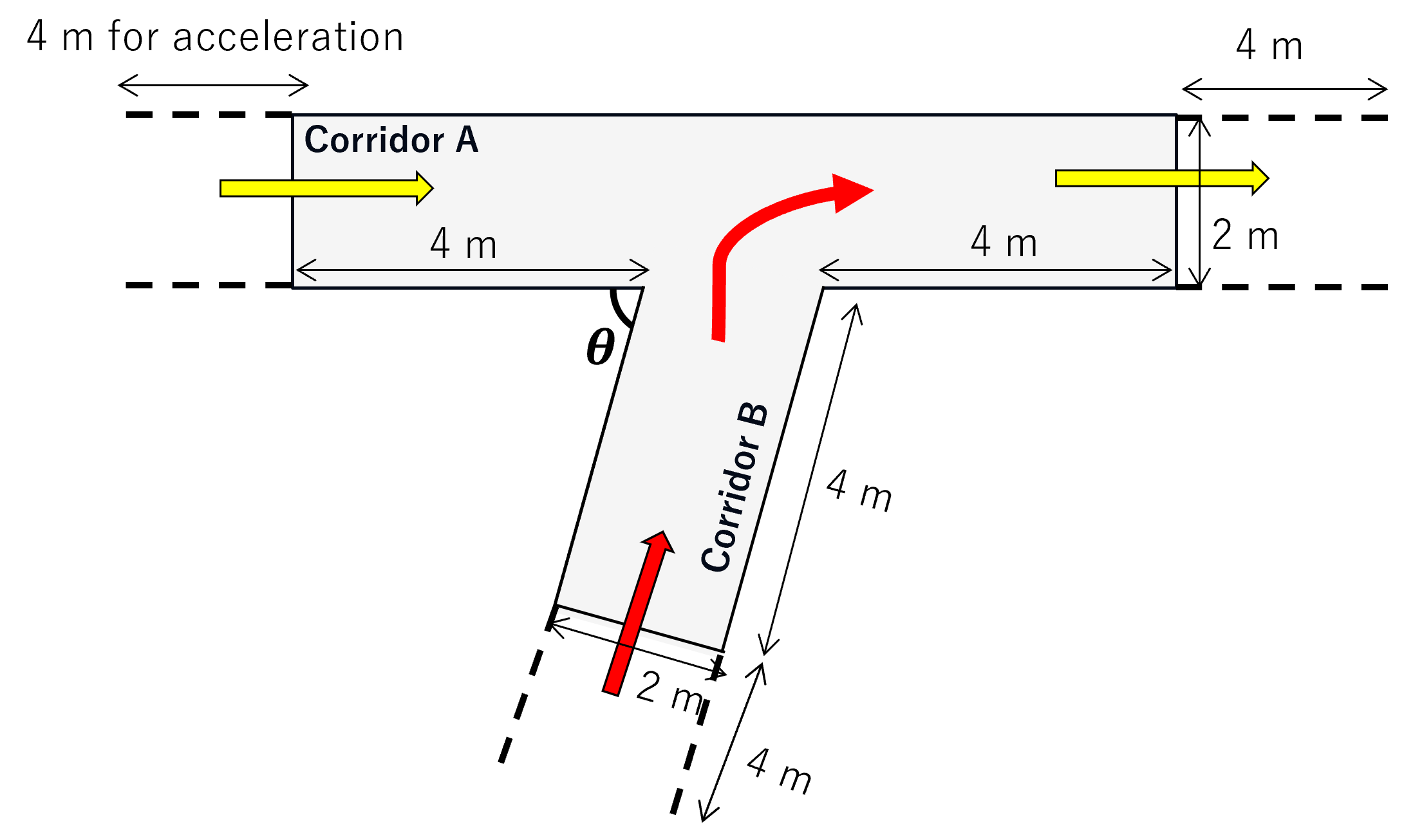}
\caption{\textbf{Experiment setup of merging T corridor.} Two pedestrian streams enter the corridor from separate branches and merge at the junction. Pedestrians in Corridor A walk straight into the merging point, while pedestrians in Corridor B approach the intersection and turn by a prescribed angle before joining the main stream. Each branch includes a 4 m acceleration zone to ensure stable entry speeds. After the merging point, both groups share the same downstream corridor. Finally, pedestrians walk through an additional 4 m downstream segment to avoid premature deceleration.\label{setupT}}
\end{figure}

The layout of T corridor  (FIG.~\ref{setupT}) consists of two pedestrian streams entered from separate branches and merged at the corner: the yellow arrow group proceeded straight along Corridor A, while the red arrow group approached along Corridor B and turned, analogous to the L corridor case. After the corner, the two streams shared the same downstream corridor.

In the T-shaped configuration, Corridor A was 4 m long, with an additional 4 m acceleration zone at its entrance to allow participants to reach a steady walking speed before entering the measurement area.
Based on these variables, 21 experimental conditions were formulated. To minimize random errors, each experimental condition was repeated several times, totaling 122 trials.

Each participant was assigned a numerical ID and directed into designated waiting areas accordingly. Upon receiving a start signal, participants simultaneously entered an acceleration area and then proceeded into the corridors. The experiment ended when the last participant exited the corridor. Participants were instructed to walk naturally, replicating their daily walking behavior.


Based on the experimental variables considered, a total of 122 trials were conducted. The specific details and configurations for each trial are summarized in Table~\ref{tab:trial_config}.

The table for the T corridor experiments follows the same format as that for the L corridor.
\begin{table}[h]
  \centering
  \caption{Summary of all T corridor experimental configurations and trial parameters.}
  \label{tab:trial_config}
  \begin{tabular}{lccccc}
    \hline
    \textbf{Angle} & \textbf{Ped} & $\boldsymbol{\rho}$ &  \textbf{Rep} & \textbf{Turning Direction} \\
    \hline
    30$^\circ$ & 4  & 0.25 & 7 & Left  \\
    30$^\circ$ & 12 & 0.75 & 6 & Left  \\
    30$^\circ$ & 24 & 1.5  & 5 & Left  \\
    30$^\circ$ & 40 & 2.5  & 5 & Left  \\
    60$^\circ$ & 4  & 0.25 & 6 & Left  \\
    60$^\circ$ & 12 & 0.75 & 5 & Left  \\
    60$^\circ$ & 24 & 1.5  & 5 & Left  \\
    60$^\circ$ & 40 & 2.5  & 3 & Left  \\
    90$^\circ$ & 4  & 0.25 & 8 & Left  \\
    90$^\circ$ & 12 & 0.75 & 6 & Left  \\
    90$^\circ$ & 24 & 1.5  & 5 & Left  \\
    90$^\circ$ & 40 & 2.5  & 4 & Left  \\
    120$^\circ$ & 4  & 0.25 & 10 & Right \\
    120$^\circ$ & 12 & 0.75 & 7  & Right \\
    120$^\circ$ & 24 & 1.5  & 6  & Right \\
    120$^\circ$ & 40 & 2.5  & 4  & Right \\
    150$^\circ$ & 4  & 0.25 & 10 & Right \\
    150$^\circ$ & 12 & 0.75 & 7 & Right \\
    150$^\circ$ & 24 & 1.5  & 6 & Right \\
    150$^\circ$ & 40 & 2.5  & 7 & Right \\
    \hline
  \end{tabular}
\end{table}
\FloatBarrier

\section{Result and discussion}\label{sec:pf}
The results are structured to identify what is unique to merging beyond pure turning.
We first compare pedestrian dynamics in L and T corridor configurations to reveal where and how merging departs from pure turning behavior.
Building on these observations, we then focus on the merging case to examine how the proposed indicators capture features arising from pedestrian interactions.
Finally, we investigate how pedestrian dynamics differ across geometric configurations based on these features.

\subsection{Comparison between L and T corridors }
It is important to compare the T corridor (turning with merging) with the L corridor (turning without merging) in order to clarify how turning alone contrasts with the combined effects of turning and merging, as previously reported~\cite{Kayvan2015}.
By comparing~FIG.\ref{setupL} and FIG.\ref{setupT}, the two geometries, it becomes possible to identify which features of the fluctuation patterns are attributable to geometric constraints and which are driven by interpersonal interactions.

This study develops a unified method for generating spatial heatmaps based on Voronoi diagrams representing individual space occupancy. Considering the discrete nature of pedestrian positions across different time frames and experimental trials, this method introduces a fixed-resolution two-dimensional spatial grid to project variable values from multiple time steps and repeated experiments at the same physical location onto a static spatial framework.

Specifically, the rectangular boundary encompassing the entire area traversed by pedestrians in all experiments is first defined and divided into uniform grid cells with a side length of 0.2 meters. The center of each grid cell serves as the spatial reference point for carrying behavioral variables. For each frame of data, the Voronoi region corresponding to each individual is extracted along with their behavioral metrics. The method then determines whether the center of each grid cell falls within a Voronoi polygon. If a grid cell is determined to be within a pedestrian’s Voronoi region in a given frame, the corresponding variable value for that individual is accumulated in the grid cell, and a weight count is recorded.

This procedure is repeated across all time frames and experimental trials, ensuring that the value assigned to each grid cell in the resulting heatmap reflects the aggregated statistics across the entire spatiotemporal dataset. Finally, the accumulated values are normalized by their respective counts to obtain the average value of each variable per grid cell over the full experimental duration. Additionally, grid cells that were never occupied by any pedestrian throughout the experiments are assigned a value of zero in order to exclude irrelevant areas and prevent misleading visual representations.

\begin{figure*}
\centering
\includegraphics[width=0.95\textwidth]{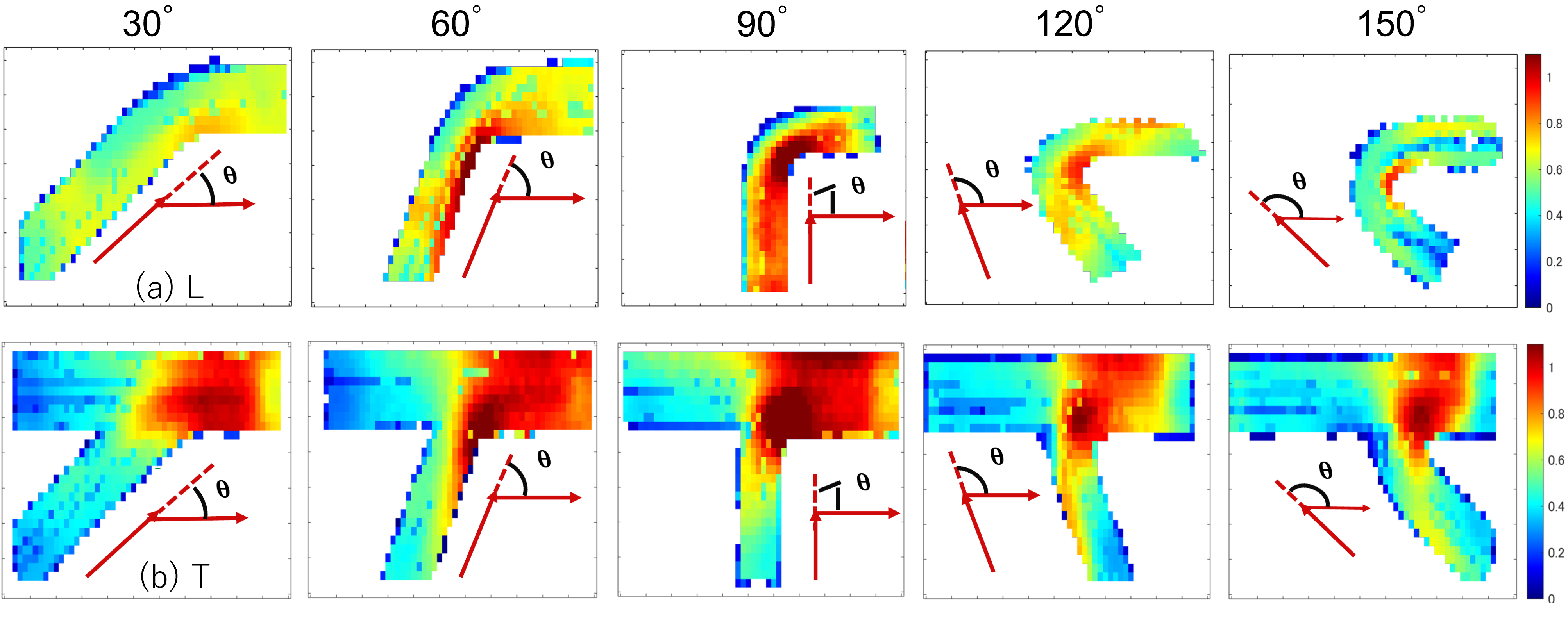}
\caption{\textbf{Spatial distributions of time-averaged Voronoi density in the (a) L corridor for 24-pedestrian experiments. and (b) T corridor for five turning angles for 40-pedestrian experiments.}
From left to right, panels correspond to turning angles of 30°, 60°, 90°, 120°, and 150°. 
Colorbars indicate the magnitude of the Voronoi density, with values ranging from low (blue) to high (red). 
In each panel, red arrows denote the walking directions of the pedestrian stream(s), and the symbol $\theta$ marks the prescribed turning angle of the corridor geometry. 
All maps are displayed on a fixed spatial grid covering the measurement region.\label{fig:voronoiLT}}
\end{figure*}

To illustrate how geometry influences crowding patterns, we investigate the spatial distributions of Voronoi density in the L and T corridor configurations, as shown in FIG.~\ref{fig:voronoiLT}. The L condition uses data from 24 participants to match the effective pedestrian count in the T condition, which involves two intersecting streams of 20 participants each.

To facilitate a direct comparison with the T corridor cases, the L corridor heatmaps for 90°, 120°, and 150° were reoriented to match the reference walking direction. For these cases, the reorientation involves a rotation of the gridded field that is not a simple index permutation in the displayed coordinate frame, so the visualization retains the full rotated domain using a loose bounding box. As a result, additional blank margins (particularly above the corridor shape) appear because the rotated effective region no longer aligns with the rectangular plotting window. These blank areas indicate padded regions outside the valid grid after rotation, not a physical change in the data magnitude or scale.

In the L corridor, FIG.~\ref{fig:voronoiLT}(a) shows that density consistently concentrates along the inner side of the bend for all turning angles, consistent with shortest-path tendencies. 
In contrast, in the T corridor, the Voronoi density increases not only during merging but also after the merging process. In addition, the high-density region shifts outward as $\theta$ increases (FIG.~\ref{fig:voronoiLT}(b)).

This shift suggests that when merging, pedestrians increasingly prioritize smoother trajectories with wider turning radii and enhanced visibility, even at the cost of longer path length, to mitigate anticipated conflicts.

\begin{figure*}
\centering
\includegraphics[width=0.95\textwidth]{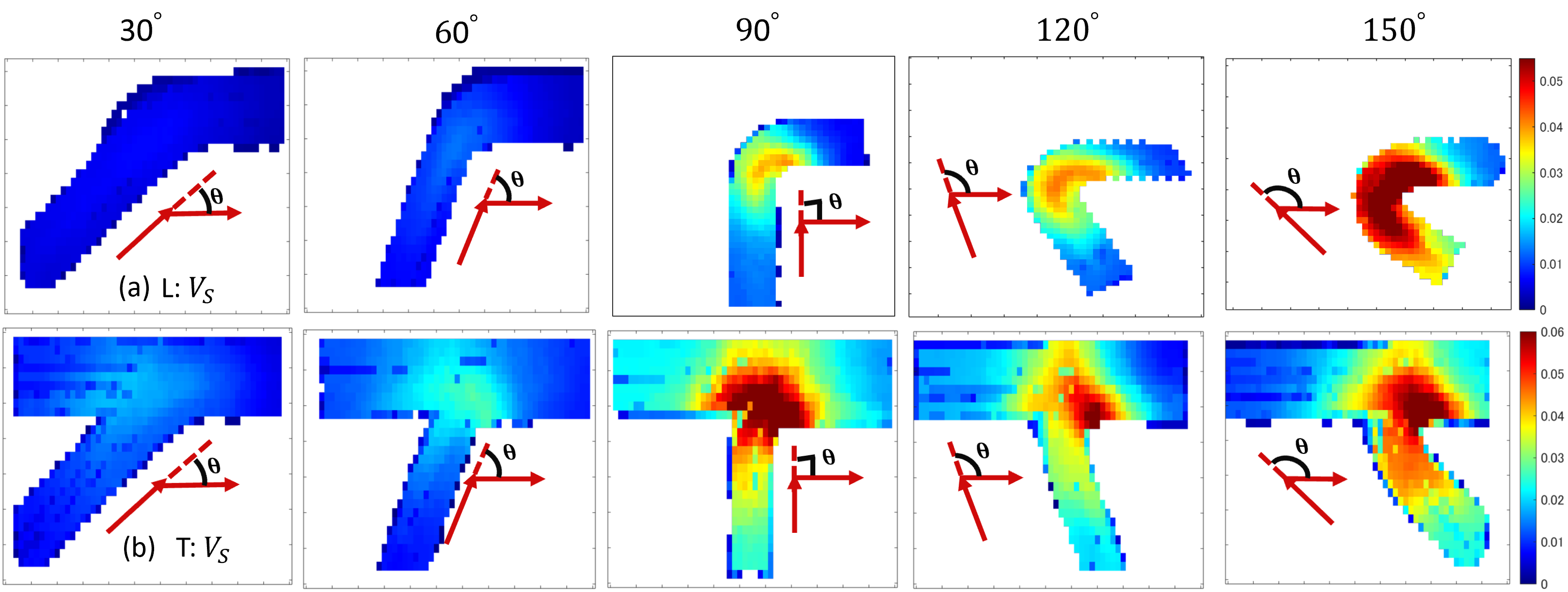}
\caption{\textbf{Spatial distributions of speed variance $V_s$ in the (a) L corridor and (b) T corridor for five turning angles for 40-pedestrian experiments.}
Panels from left to right correspond to turning angles of 30°, 60°, 90°, 120°, and 150°.
Colorbars indicate the magnitude of the speed variance $V_s$, with values displayed from low (blue) to high (red).
In each panel, red arrows denote the walking directions of the pedestrian stream(s), and the symbol $\theta$ marks the prescribed turning angle of the corridor geometry.
All maps are shown on a fixed spatial grid covering the measurement region, and variance values are averaged over all frames and repeated trials under the same experimental condition.\label{fig:vsLT}}
\end{figure*}

We next examine how geometric configuration influences the smoothness of pedestrian movement. To ensure a consistent basis for comparison and to focus on behavior at the merging region, all subsequent analyses of both the L and T configurations are conducted using experiments with 40 pedestrians.
For this purpose, we analyze the spatial distribution of the speed variance $V_s$, which captures local fluctuations arising from interactions among pedestrians. 
FIG.~\ref{fig:vsLT} presents the corresponding spatial distributions of $V_s$ across all turning angles.

In FIG.~\ref{fig:vsLT}(a), L corridor, high $V_s$ appears in a symmetric lobe around the corner apex, increasing with turning angles. This pattern reflects transient turning-induced adjustments (deceleration, re-acceleration, and local evasions) concentrated near the bend.

On the other hand, in FIG.~\ref{fig:vsLT}(b) T corridor, the dominant hotspot is found downstream of the corner at the merger spot, rather than at the corner itself. $V_s$increases sharply where the turning stream meets the straight stream, indicating interaction-driven fluctuations.

Although elevated $V_s$ values are observed downstream of the corner in the L corridor at
large angles (e.g., 150°), similar high values are also present upstream, forming a nearly
symmetric distribution around the corner. This pattern indicates that the fluctuations are
turning-driven rather than interaction-driven effects, as sharp turns require pedestrians to adjust their motion over a longer spatial range.

\begin{figure*}
\centering
\includegraphics[width=0.95\textwidth]{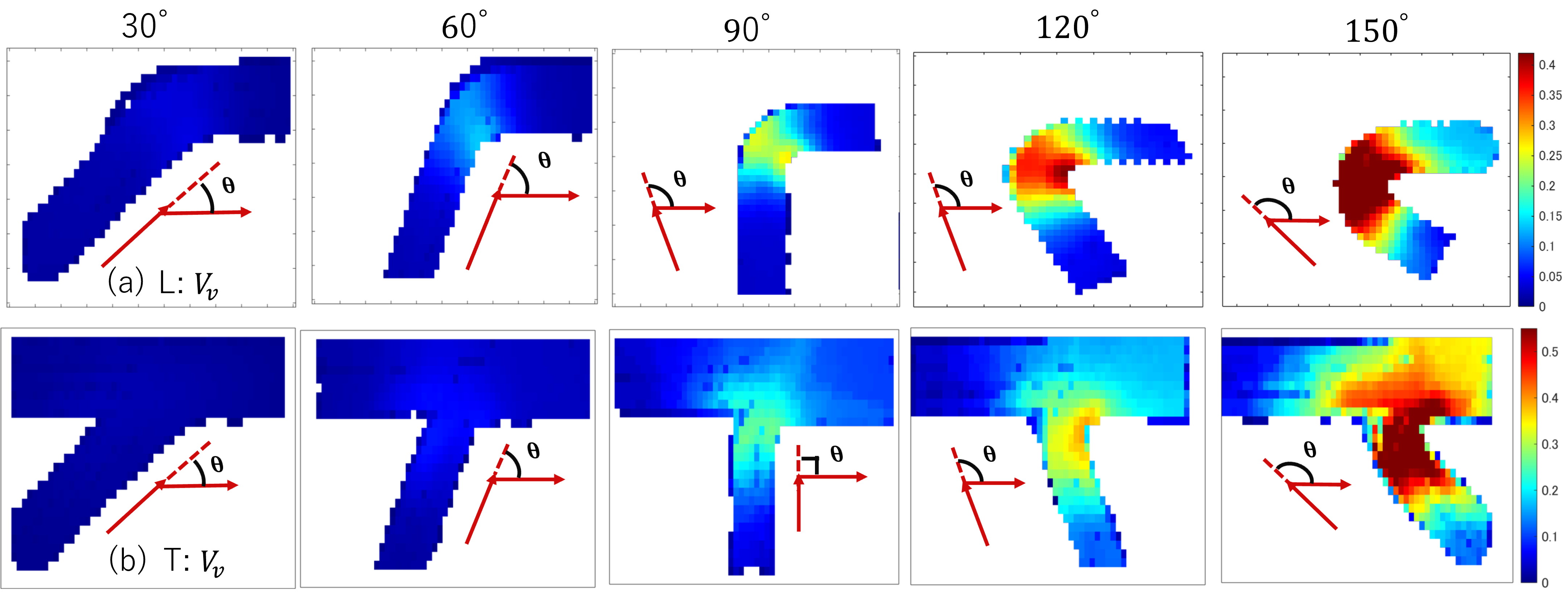}
\caption{\textbf{Spatial distributions of velocity variance $V_v$ in the (a) L corridor and (b) T corridor for five turning angles for 40-pedestrian experiments.}
Panels from left to right correspond to turning angles of 30°, 60°, 90°, 120°, and 150°.
Colorbars represent the magnitude of the velocity variance $V_v$, with lower values shown in blue and higher values in red.
Red arrows in each panel indicate the walking directions of the pedestrian stream(s), and the symbol $\theta$ marks the prescribed turning angle of the corridor geometry.
All distributions are displayed on a fixed spatial grid covering the experimental measurement region, and values are averaged over all frames and repeated trials under the same experimental condition.\label{fig:vvLT}}
\end{figure*}

In addition to the speed variance, it is also necessary to examine the variance of the velocity vectors, which reflect fluctuations in both the magnitude and the direction of pedestrian motion. This measure complements speed variance by capturing not only fluctuations in walking speed but also variations in walking direction, allowing us to assess how geometric configuration affects the directional consistency of pedestrian movement. FIG.~\ref{fig:vvLT} shows the spatial distributions of the velocity variance $V_v$ for all turning angles.

In the L corridor, the highest velocity variance values are concentrated within the turning segment itself (FIG.~\ref{fig:vvLT}(a)). As pedestrians traverse the corner, the variance increases markedly, forming a localized region of elevated fluctuation along the inner side of the bend. This pattern indicates that direction adjustments are primarily executed during the act of turning.

In the T corridor, however, the location of increased velocity variance is shifted upstream of the turning point (FIG.~\ref{fig:vvLT}(b)). A pronounced high velocity variance area appears before pedestrians reach the junction, revealing that direction adjustments begin prior to entering the curved section. This upstream elevation reflects an anticipatory modification of movement, consistent with the need to accommodate the forthcoming merging interaction. Downstream of the junction, velocity variance remains visible but becomes more spatially dispersed, reflecting the additional adjustments made as the two streams reorganize into a single flow.

In summary, merging can be characterized by the speed variance and the velocity variance rather than by the Voronoi density.
\FloatBarrier

\subsection{Spatial Distribution of T corridor}
Building on the comparison with the L corridor, we have shown that speed or velocity variance can characterize merging behavior even under different geometric configurations. This motivates a focused analysis of the T corridor to clarify which specific features of merging dynamics are captured by the variance-based indicator and to distinguish the physical mechanisms reflected in scalar ($V_s$) versus vector ($V_v$) measures.
This section aims to assign physically meaningful characteristics to merging behavior by examining how the proposed indicators reflect interaction-driven fluctuations. 
\begin{figure*}
\centering
\includegraphics[width=0.95\textwidth]{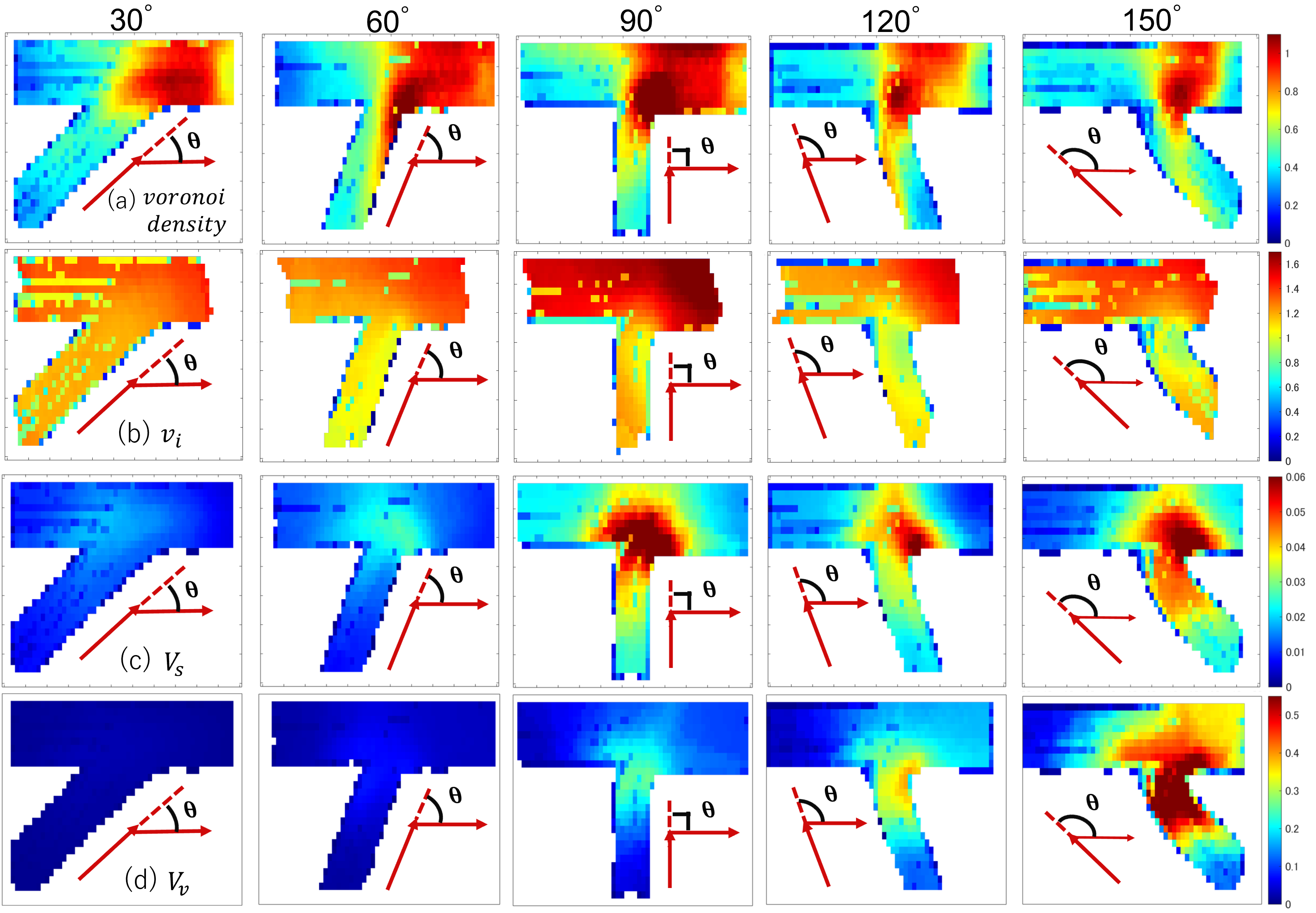}
\caption{\textbf{Spatial distributions of (a)Voronoi density, (b)individual speed, and (c)speed variance $V_s$ (d)velocity variance $V_v$, adopted from FIG.~\ref{fig:voronoiLT}--FIG.~\ref{fig:vvLT}}.
Panels from left to right correspond to turning angles of 30°, 60°, 90°, 120°, and 150°.
In all panels, colorbars indicate the numerical magnitude of each variable, with warmer colors representing higher values and cooler colors representing lower values. The corridor boundaries are outlined for reference, and arrows denote the primary walking directions of the two pedestrian streams. These maps are obtained by aggregating data over all frames and trials under the same 40-pedestrian experimental condition.\label{fig:heatmap1_1}}
\end{figure*}

From the perspective of spatial distribution, the Voronoi density patterns exhibit clear variations across different angles. As shown in FIG.~\ref{fig:heatmap1_1}(a), high-density areas occur mainly in the main passage after merging, particularly near the corners. This indicates that pedestrians prefer shorter paths closer to the inner side when turning, resulting in increased space usage at corners.

As discussed in the previous subsection, the Voronoi density increases not only at the moment of merging but also persists after merging. This effect is particularly evident near the corners, where pedestrians tend to choose shorter paths.

Regarding individual speed, FIG.~\ref{fig:heatmap1_1}(b) demonstrates that pedestrians on the straight main path (Corridor A) generally maintain higher speeds compared to those on the branch path (Corridor B), suggesting that merging and turning decrease walking speed, especially under conditions with larger angles. Notably, apart from the 30° scenario, pedestrians moving straight consistently maintain higher speeds, while those turning pedestrians on Corridor B maintain notably lower speeds. This difference in speed arises not only from geometric constraints but also from differences in visibility and behavioral strategies before merging. This is mainly due to the following reason: Pedestrians who turn have a clear view of straight-moving pedestrians ahead and therefore proactively reduce their speed in anticipation of merging with them. In contrast, straight-moving pedestrians (except in the 30° scenario) typically cannot easily observe pedestrians approaching from the side paths and thus maintain their original speeds until merging occurs. This ''primary/secondary'' behavioral dynamic between pathways creates structural differences in walking speeds.

Additionally, FIG.~\ref{fig:heatmap1_1}(b) shows that the highest speeds are not located in areas with the lowest density but rather in the main passage after merging. This indicates that increased density does not necessarily lead to reduced speed, as there can be coordinated movement characterized by ``high density and high speed'' flow, although speed typically decreases as density increases. 

Speed and density are standard indicators of risk, but under these merging conditions, relying solely on density or speed individually is insufficient. In complex merging scenarios, pedestrian instability primarily arises from behavioral and interactional factors such as anticipation, hesitation, and oscillatory movement.

Therefore, density and speed-based measures alone provide only partial descriptions of pedestrian flow. As a result, the dynamical manifestations of pedestrian interactions, such as local adjustments and coordination processes, become essential for understanding the flow. In merging scenarios, where interactions are highly localized around the merging point, these limitations become especially pronounced.

As shown in FIG.~\ref{fig:heatmap1_1}(c), the distribution of speed variance $V_s$ consistently highlights the region immediately after the merging point, particularly where pedestrians from the side corridor complete the turn and enter the main stream. This post-merging zone exhibits elevated $V_s$ values under all turning angle conditions, forming an elongated high variance band along the downstream flow. The pattern reflects significant fluctuations in walking speed due to directional alignment difficulties and mutual avoidance behaviors that arise as pedestrians with different trajectories and speeds are brought together by the merging geometry, leading to locally elevated densities. In essence, high $V_s$ values reveal the presence of instability caused by inter-personal interactions during merging, where individuals struggle to maintain uniform movement in the face of geometric constraints and interaction pressure.

As already discussed in the previous subsection, the region where the speed variance $V_s$ is high corresponds to the merging region, where pedestrians from the side corridor complete the turn and enter the main flow (FIG.~\ref{fig:heatmap1_1}(c)). This high-speed variance region is elongated toward the downstream direction. This reflects the fact that pedestrians slow down to align their walking direction with that of other groups and avoid conflicts during merging. In this sense, a high-speed variance corresponds to interpersonal interactions.

In contrast, FIG.~\ref{fig:heatmap1_1}(d) shows that velocity variance $V_v$ reaches its maximum upstream of the turning area, rather than in the merged zone. These high $V_v$ regions are especially prominent in cases with larger turning angles and consistently precede the point of intersection. This spatial shift suggests that directional variation begins before pedestrians enter the corner, implying that individuals engage in anticipatory adjustments to their walking direction. Such pre-turning directional changes are likely driven not only by geometric constraints but also by an awareness of upcoming interactions, as pedestrians adapt early to avoid future conflicts. While $V_s$ captures instability arising from direct interactions in the merged flow, $V_v$ reflects preparatory phase of path negotiation under complex spatial conditions.

In contrast, the velocity variance $V_v$ reaches its maximum upstream of the turning area, rather than in the merged zone (FIG.~\ref{fig:heatmap1_1}(d)). It increases with increasing turning angle. These results suggest that pedestrians from the side corridor adjust their direction of motion before entering the corner, as discussed previously.

Therefore, speed variance $V_s$ is more suitable for identifying instability arising from conflict avoidance between pedestrians, reflecting coordination-related speed fluctuations, while velocity variance $V_v$ effectively captures instability caused by external geometric changes, highlighting directional adjustments due to path geometry. Together, these two indicators comprehensively characterize pedestrian dynamics and potential risk areas within complex merging scenarios.

\begin{figure*}
\centering
\includegraphics[width=0.95\textwidth]{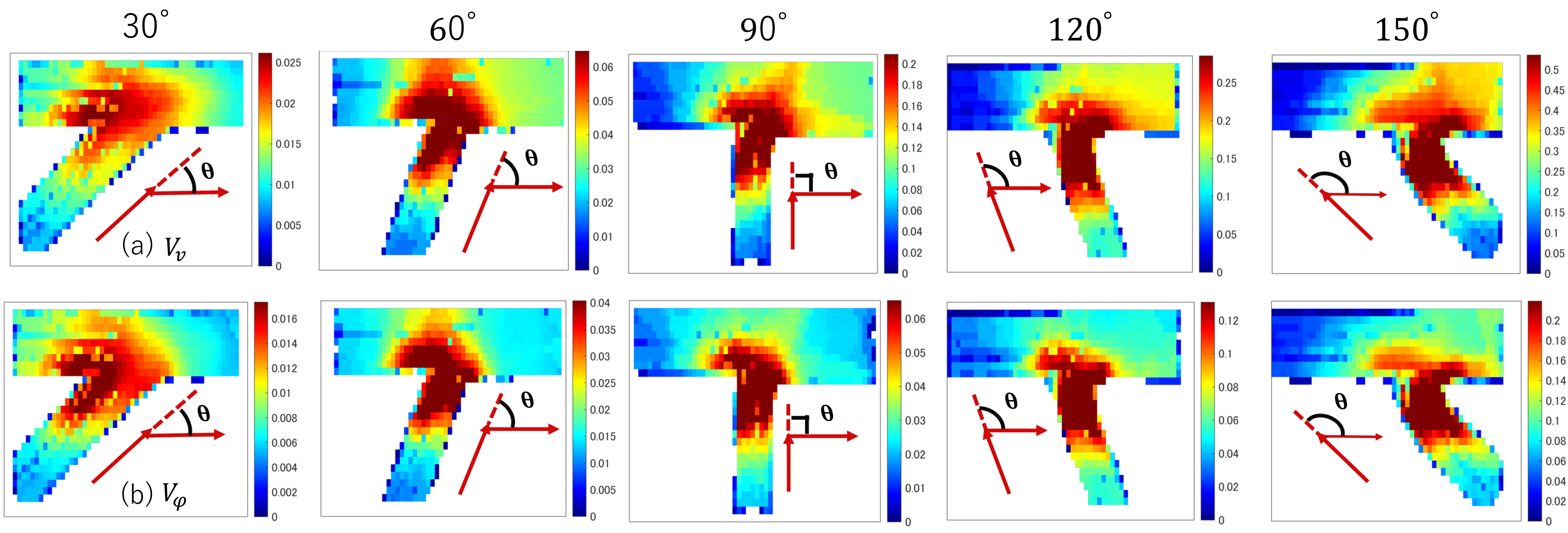}
\caption{\textbf{Spatial distributions of velocity variance $V_v$ and directional variance $V_\phi$ in the T corridor.}
Panel (a) shows the spatial distribution of the velocity variance $V_v$, computed from deviations of pedestrian velocity vectors within each Voronoi neighborhood and mapped onto a fixed spatial grid.
Panel (b) presents the spatial distribution of the directional variance $V_\phi$, evaluated from the circular variance of walking directions.
Panels from left to right correspond to turning angles of 30°, 60°, 90°, 120°, and 150°.
In both panels, colorbars represent the numerical magnitude of each variance measure, with warmer colors indicating higher values. 
Corridor boundaries are outlined for reference, and arrows indicate the primary walking directions of the two pedestrian streams.
All values are obtained by aggregating measurements over all frames and repeated trials under the same 40-pedestrian experimental condition.
\label{fig:vv_vtheta}}
\end{figure*}

To further verify whether the variance of the velocity vectors truly reflects the turning behavior, we additionally computed the variance of the pedestrians’ turning direction $V_\phi$,

When comparing the spatial distribution of the direction variance $V_\phi$ with that of the velocity variance $V_v$, the two fields exhibit highly similar patterns across all geometries (FIG.~\ref{fig:vv_vtheta}). This consistency indicates that the velocity variance is primarily governed by the rotation of the pedestrians’ walking direction.

Taken together, these results indicate that the velocity variance is primarily governed by the rotation of pedestrians’ walking direction, whereas the disorder arising from pedestrian–pedestrian interactions is more directly captured by the speed variance. Therefore, $V_s$ is adopted as the primary indicator in this study to characterize interaction-driven disorder in merging flows.

\FloatBarrier

\subsection{Angle dependency}
In view of the above results, all subsequent analyses of angle dependency are performed using the speed variance $V_s$.

\begin{figure}
\centering
\includegraphics[width=1\textwidth]{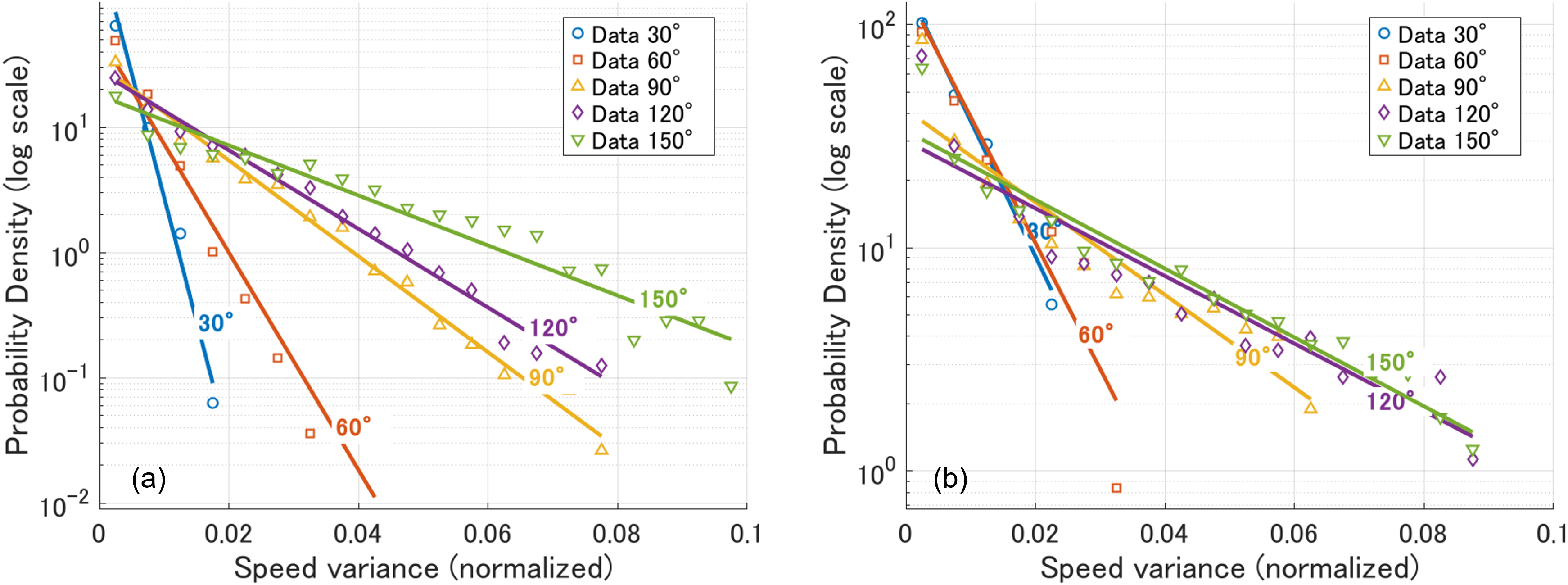}
\caption{\textbf{Probability density functions of normalized speed variance in the (a) L corridor, and (b) T corridor for five turning angles.}
Scatter points show the empirical probability density of the normalized speed variance for each angle condition (30°, 60°, 90°, 120°, 150°).
Solid lines represent the corresponding fitted curves plotted in the same color as the data points.
The vertical axis is shown on a logarithmic scale, and the horizontal axis indicates the normalized speed variance.
All datasets are computed from aggregated frames and repeated trials under each angle condition.\label{PDFLT}}
\end{figure}

To further investigate how the turning angle influences the statistical structure of local fluctuations, we examine the probability density functions (PDFs) of the speed variance $V_s$. While the spatial heatmaps reveal where fluctuations occur, the PDFs provide a complementary description by characterizing the overall distribution of fluctuation magnitudes within each geometric condition.
The PDFs were obtained through the following procedure. First, all $V_s$ samples from multiple experimental runs belonging to the same angle were pooled into a single dataset. Outliers were then removed, and the remaining values were grouped into fixed-width bins to estimate their probability density. This procedure produces a stable and comparable statistical representation of the fluctuation distribution for each angle.
FIG.~\ref{PDFLT} shows the resulting PDFs of the normalized speed variance $V_s$ for the five turning angles. Normalization was applied so that datasets from different geometries fall within a comparable numerical range.

In the FIG.~\ref{PDFLT}, each point corresponds to the center of a bin obtained from the PDF calculation, while the solid lines represent exponential fits drawn in the logarithmic vertical axis. This representation allows direct visual comparison of how the tail behavior of the fluctuation distribution varies with turning angle.

The fitted curves reveal a distinct dependence on the turning angle. For 30° and 60°, the PDFs decrease rapidly, indicating that large fluctuations in $V_s$ occur with very low probability. For angles of 90° and above, the decline becomes more gradual, and the fitted curves for 90°, 120°, and 150° exhibit similar slopes. This transition suggests that the statistical distribution of fluctuation intensity changes with the turning angle, and that sharper turns are associated with slower decay in the distribution.

A closer inspection reveals differences between the L corridor and the T corridor. In the L corridor (FIG.~\ref{PDFLT}(a)), the rate of decay becomes progressively slower with increasing angle, and the relationship between angle and slope is approximately monotonic. 

In the T corridor, as shown in FIG.~\ref{PDFLT}(b), however, the trend is not strictly monotonic: the 120° and  150° cases show the slowest decay, resulting in the broadest distribution among all tested angles. In both configurations, the 90° case marks a turning point, where the PDFs begin to deviate from the rapid-decay regime seen at small angles. This may correspond to a regime in which pedestrian motion becomes less smooth, giving rise to more pronounced speed fluctuations.

\begin{figure*}
\centering
\includegraphics[width=0.95\textwidth]{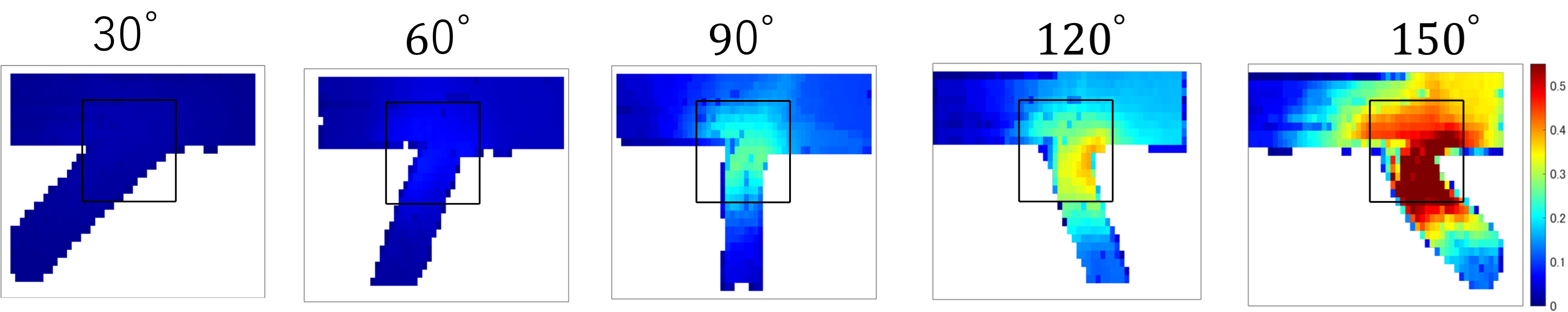}
\caption{\textbf{Extracted center region (black square) for different turning angles.}
Panels correspond to turning angles of 30°, 60°, 90°, 120°, and 150° (from left to right).
Each panel shows the spatial map of the measured variable on a fixed grid within the T-shaped corridor, and the black square indicates the predefined center region used for subsequent quantitative analysis.\label{center}}
\end{figure*}

To focus on the combined effects of turning and merging, a center region was defined within each scenario, as shown in FIG.~\ref{center}. This rectangular area (4~m in length and 2.4~m in width) captures the spatial zone where two key processes occur in sequence: pedestrians complete the turning motion and immediately interact with the opposing flow. Within this region, both geometric constraints and interpersonal interference are simultaneously at play. In the following analysis, we evaluate the average values of key indicators in this center region to compare stability across different turning angles and corridor geometries.

\begin{figure}
\centering
\includegraphics[width=1\textwidth]{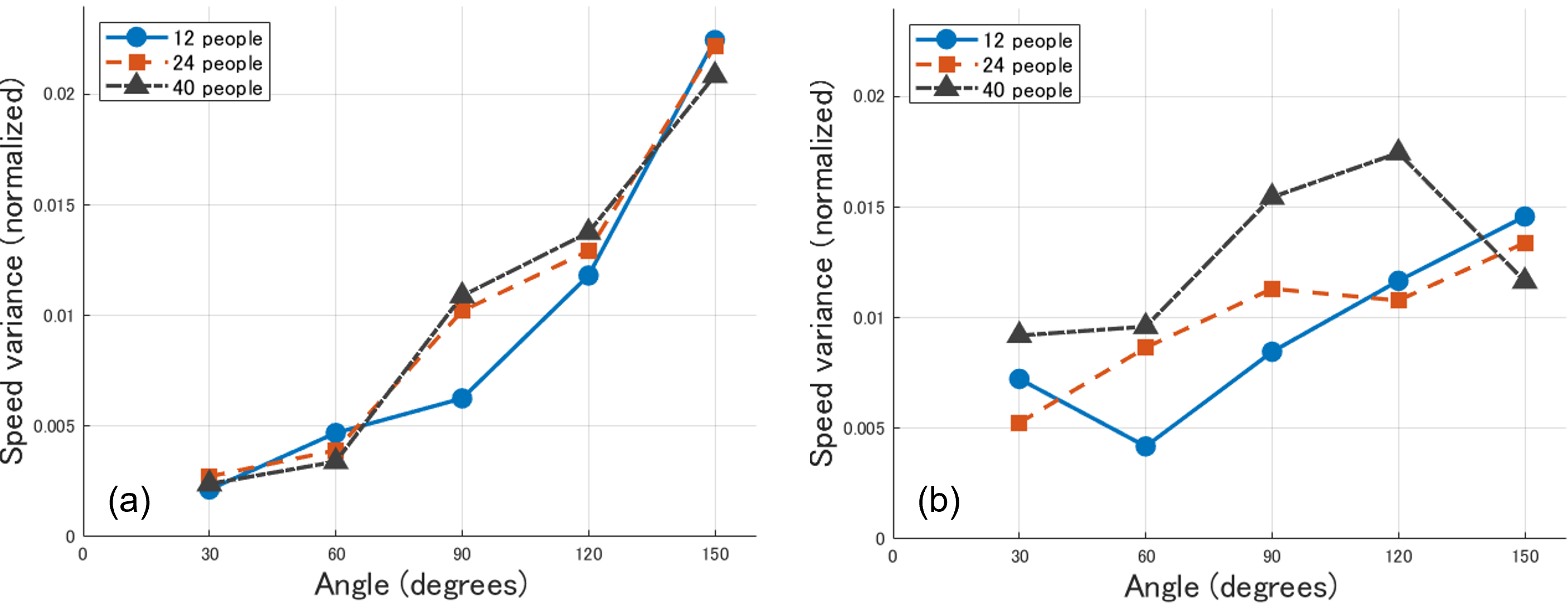}
\caption{\textbf{Area-averaged normalized speed variance in the(a)L corridor, and (b)T corridor for different turning angles and population sizes.}
The plotted curves correspond to experiments with 12, 24, and 40 participants.
The horizontal axis shows the prescribed turning angle, and the vertical axis indicates the area-averaged normalized speed variance computed within the predefined center region.
Markers and colors distinguish the three populations, and each value represents the average over all frames and repeated trials for the corresponding experimental condition.\label{Angle_Vs}}
\end{figure}

First, as shown in FIG.~\ref{Angle_Vs}(a), in the L corridor, a clear population effect emerges most prominently at the 90° condition: the mean normalized values of $V_s$ separate distinctly 12 participants from 24 and 40 participants, indicating that fluctuations at this angle are particularly sensitive to density. In addition, the overall trend shows a monotonic increase in $V_s$  with angle, suggesting that, in the absence of merging, the variability in speed is primarily governed by turning angle: sharper turns lead to larger adjustments, and the effect scales in a nearly proportional manner with the turning angle.
In the T corridor (FIG.~\ref{Angle_Vs}(b)), the pattern differs. Except for the extreme cases of 30° and 150°,  $V_s$ tends to rise with increasing participant number, implying that interactions introduced by merging become stronger as density increases. The angle dependence also deviates from the L corridor scenario: rather than increasing proportionally with the turning angle, $V_s$ begins to rise gradually around 90° and reaches its highest values near 120° for 40 people. Beyond 90°, the influence of geometry becomes less dominant, and the variability in speed appears increasingly shaped by the local interactions inherent to the merging process.

\begin{figure*}
\centering
\includegraphics[width=1\textwidth]{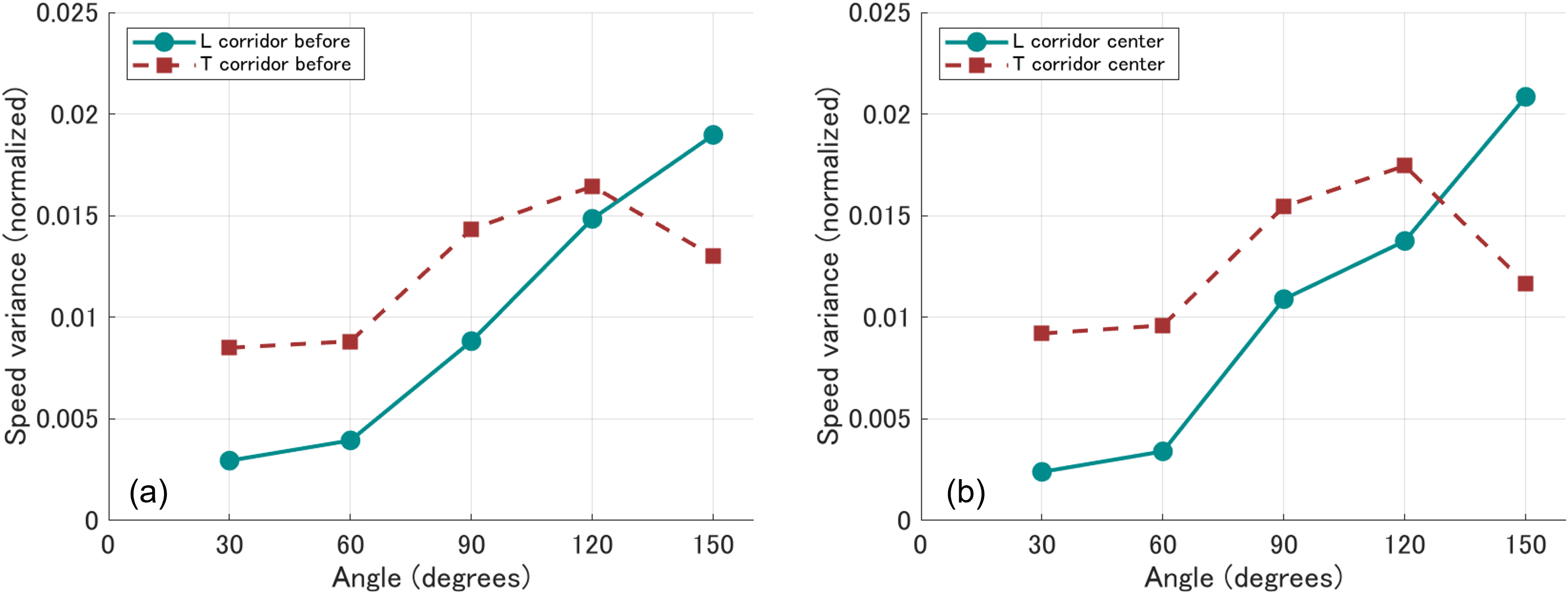}
\caption{\textbf{Area-averaged normalized speed variance in the (a)before turning region, and (b)turning center region for the L and T corridors across different turning angles for 40-pedestrian experiments.}
Curves represent the normalized speed variance computed in the center region, shown separately for the L corridor (green markers) and the T corridor (red markers).
The horizontal axis indicates the prescribed turning angle, and the vertical axis shows the corresponding area-averaged normalized speed variance.
All values are averaged over all frames and repeated trials under each experimental condition\label{LT_Vs}}
\end{figure*}

To verify whether pedestrians indeed begin adjusting their motion in advance, and to identify the turning-angle range at which interaction-driven adjustments become more influential than the geometric constraint itself, we compare the fluctuation levels across different regions and geometries.
FIG.~\ref{LT_Vs}(a) and FIG.~\ref{LT_Vs}(b) present the overall trend of how the mean normalized speed variance changes with turning angle in the 40-person experiments. It then compares the values obtained in the L and T corridors, evaluated separately in the pre-turning region and the turning-center region.

T across almost all angles, the T corridor exhibits higher $V_s$ values than the L corridor, except for the 150° condition. This indicates that the presence of merging consistently increases fluctuation levels beyond those produced by turning geometry alone.
In the L corridor, $V_s$ increases monotonically with the turning angle in both regions, showing that the fluctuation level is primarily governed by the geometric sharpness of the turn. 
In contrast, the T corridor displays a combined influence of geometry and merging interactions. For angles below 90°, $V_s$ in the T corridor remains clearly higher than in the L corridor, suggesting that merging effects amplify the fluctuations induced by turning.

A notable change appears at 120°. At this angle, the difference between the L and T corridor values decreases substantially, particularly in the pre-turning region. This reduction suggests that pedestrians begin adjusting their motion before reaching the merging point, and such anticipatory adjustments lessen the direct influence of geometric turning constraints. As a result, the increase in $V_s$ that would normally accompany larger turning angles is partially suppressed once merging becomes the dominant consideration.

At 150°, the relationship reverses: the T corridor values fall below those of the L corridor in both regions. This indicates that at very large angles, the geometric effect that produces strong fluctuations in the L corridor is suppressed in the T corridor by interactions associated with merging. As pedestrians adapt to the increased interaction area, the influence of the turning angle becomes less pronounced. 

The above analyses indicate that the smoothness of pedestrian flow is governed by two principal components:  
(i) the effect of pedestrian–pedestrian interactions, particularly those associated with merging, and  
(ii) the effect of the variation in the directions of motion at the corner.

To formalize this idea, we introduce two functions, $E(I;\theta)$ and $E(G;\theta)$, which represent the interaction effect and the geometrical effect on the fluctuation level at turning angle $\theta$.

In the L corridor, where no merging occurs, the observed fluctuation can be attributed solely to the geometric configuration of the turn. The total effect on smoothness is therefore written as  
\begin{equation}
    E_L(\theta) = E(G;\theta).
    \label{eq:EL}
\end{equation}

In the T-shaped corridor, both interaction and geometry contribute. We express the total effect as  
\begin{equation}
    E_T(\theta) = E(I;\theta) + \left(1 + \beta(\theta)\right) E(G;\theta),
    \label{eq:ET}
\end{equation}
where $\beta(\theta)$ is a dimensionless combined coefficient that modulates the geometric contribution in the presence of merging.
When $\beta(\theta) > 0$, the geometric effect is strengthened by interactions;  
when $\beta(\theta) \in [-1, 0]$, the geometric effect is partially suppressed;  
and when $\beta(\theta) < -1$, the interaction term overcompensates the geometric contribution so strongly that the net influence of geometry is effectively reversed.

Qualitatively, this interpretation is consistent with the empirical trends:  
for $\theta < 90^\circ$, $\beta(\theta)$ is expected to be positive, indicating that merging amplifies the angle dependency effect observed in the L corridor.  

Around $\theta = 120^\circ$, $\beta(\theta)$ approaches zero or slightly negative values, reflecting anticipatory adjustments that weaken the direct impact of geometry.  

At $\theta = 150^\circ$, $\beta(\theta)$ becomes strongly negative, corresponding to a regime in which interaction effects suppress the geometric influence to the extent that the T corridor appears smoother than the L corridor despite the larger turning angle.
\FloatBarrier

\section{Conclusion}\label{sec:pf}
To quantify behaviors characteristic of merging, in which a pedestrian group enters another group and aligns with it, this study presented a comparative analysis of pedestrian dynamics in turning and merging corridors by examining L-shaped and T-shaped geometries under identical experimental conditions. Using Voronoi-based measures of speed variance $V_s$, velocity variance $V_v$, and directional variance $V_\phi$, we evaluated how turning angle and interpersonal interactions influence the smoothness of pedestrian flow.

The results show that turning and merging generate distinct spatial patterns of density and fluctuations. In the L corridor, high values of $V_s$ appear mainly around the corner apex and increase with the turning angle, indicating that geometric curvature governs local adjustments in speed. In the T corridor, the highest fluctuations occur downstream of the corner in the merging zone, where interactions between the two streams intensify instability. Velocity variance and directional variance also reveal characteristic differences: in the L corridor, they peak within the turning region, whereas in the T corridor, they shift upstream, demonstrating anticipatory directional changes before pedestrians reach the merging point.

Angle dependence further clarifies the different roles of geometry and interactions. In the L corridor, $V_s$ increases monotonically with the turning angle. In the T corridor, the trend is non-monotonic. Fluctuations rise near $90^\circ$, reach their highest levels around $120^\circ$, and diminish at $150^\circ$. These patterns indicate that merging interactions strengthen geometric effects at moderate angles but may weaken them at very large angles as pedestrians adapt their behavior in advance.

To interpret these findings, we introduce a conceptual decomposition that separates fluctuation levels into contributions arising from geometric constraints and from pedestrian–pedestrian interactions. Within this framework, pure turning configurations are dominated by geometry-induced adjustments associated with path curvature, whereas turning plus merging configurations are additionally shaped by interaction-induced processes such as conflict avoidance, anticipation, and local reorganization of movement. This decomposition clarifies why the two configurations exhibit distinct spatial patterns and statistical characteristics of instability. Moreover, it provides a coherent explanation for previously reported inconsistencies in the effects of turning angle, by showing that the relative importance of geometric and interaction components varies systematically with angle.

Overall, the comparative results highlight that geometry and interpersonal interactions jointly determine walking smoothness in complex corridors. The variance-based indicators provide a physically interpretable tool for identifying where instability emerges and how it evolves across angles and densities. Future work will extend this analysis to more complex layouts and diverse populations and will explore the use of these indicators for real-time monitoring and design evaluation.



\clearpage 


This work was supported by JST SPRING Grant Number JPMJSP2108, JST-Mirai Program Grant Number JPMJMI20D1, Japan, National Natural Science Foundation of China (No. 52402376), JSPS KAKENHI Grant Number JP23K21019, JP23K13521, JSPS KAKENHI (JP25K17790), the Odakyu Foundation Research Program, the Obayashi Foundation Research Program (ST), JP23K20947. In completing this manuscript, the authors used OpenAI's ChatGPT to refine the language and check for writing errors and ensure grammatical accuracy. 




\printcredits

\bibliographystyle{cas-model2-names}

\bibliography{References}

@article{Polli2023,
    title = {Analytical model for collision probability assessments with large satellite constellations},
    author = {Eduardo Maria Polli and Juan Luis Gonzalo and Camilla Colombo},
    journal = {Advances in Space Research},
    volume = {72},
    number = {7},
    pages = {2515-2534},
    year = {2023},
    note = {Space Environment Management and Space Sustainability},
    issn = {0273-1177},
    doi = {https://doi.org/10.1016/j.asr.2022.07.055},
    url = {https://www.sciencedirect.com/science/article/pii/S027311772200686X},
}

@article{Aditya2024,
    title = {A review on air traffic flow management optimization: trends, challenges, and future directions},
    author = {Verma Aditya and Dande Sureshkumar Aswin and Somasundaram Vanitha Dhaneesh and Sakthivelan Chakravarthy and Bhukya Shanmuk Kumar and Marimuthu Venkadavarahan},
    doi = {10.1007/S43621-024-00781-7},
    isbn = {0123456789},
    issn = {2662-9984},
    issue = {1},
    journal = {Discover Sustainability 2024 5:1},
    month = {12},
    pages = {519-},
    publisher = {Springer},
    volume = {5},
    url = {https://link.springer.com/article/10.1007/s43621-024-00781-7},
    year = {2024}
}

@article{Kang2022,
   title = {Study of narrow waterways congestion based on automatic identification system (AIS) data: A case study of Houston Ship Channel},
   author = {Masood Jafari Kang and Sepideh Zohoori and Maryam Hamidi and Xing Wu},
   doi = {10.1016/J.JOES.2021.10.010},
   issn = {2468-0133},
   issue = {6},
   journal = {Journal of Ocean Engineering and Science},
   month = {12},
   pages = {578-595},
   publisher = {Elsevier},
   volume = {7},
   url = {https://www.sciencedirect.com/science/article/pii/S246801332100125X?utm_source=chatgpt.com},
   year = {2022}
}

@article{Wang2025,  
   title = {Collaborative rescheduling of train timetables to relieve passenger congestions in an urban rail transit network: a rolling horizon approach},
   author = {Fangsheng Wang and Pengling Wang and Xiaoyu Hao and Rudong Yang and Ruihua Xu},
   doi = {10.1016/J.IJTST.2024.09.001},
   issn = {2046-0430},
   journal = {International Journal of Transportation Science and Technology},
   month = {9},
   pages = {193-210},
   publisher = {Elsevier},
   volume = {19},
   url = {https://www.sciencedirect.com/science/article/pii/S2046043024001035?utm_source=chatgpt.com},
   year = {2025}
}

@Article{Mushtaq2021,
    AUTHOR = {Mushtaq, Anum and Haq, Irfan ul and Nabi, Wajih un and Khan, Asifullah and Shafiq, Omair},
    TITLE = {Traffic Flow Management of Autonomous Vehicles Using Platooning and Collision Avoidance Strategies},
    JOURNAL = {Electronics},
    VOLUME = {10},
    YEAR = {2021},
    NUMBER = {10},
    ARTICLE-NUMBER = {1221},
    URL = {https://www.mdpi.com/2079-9292/10/10/1221},
    ISSN = {2079-9292},
    DOI = {10.3390/electronics10101221}
}

@article{Helbing2007,
  author = {Helbing, Dirk and Johansson, Anders and Al-Abideen, Habib Zein},
  title = {Dynamics of crowd disasters: An empirical study},
  journal = {Phys. Rev. E},
  volume = {75},
  issue = {4},
  pages = {046109},
  numpages = {7},
  year = {2007},
  month = {Apr},
  publisher = {American Physical Society},
  doi = {10.1103/PhysRevE.75.046109},
  url = {https://link.aps.org/doi/10.1103/PhysRevE.75.046109}
}

@article{Kayvan2014,
    author = {Kayvan Aghabayk and Omid Ejtemai and Majid Sarvi and Amir Sobhani},
    title = {Understanding Pedestrian Crowd Merging Behavior},
    journal = {Transportation Research Procedia},
    volume = {2},
    pages = {768-773},
    year = {2014},
    note = {The Conference on Pedestrian and Evacuation Dynamics 2014 (PED 2014), 22-24 October 2014, Delft, The Netherlands},
    issn = {2352-1465},
    doi = {https://doi.org/10.1016/j.trpro.2014.09.086},
    url ={https://www.sciencedirect.com/science/article/pii/S2352146514001227}
}

@article{Shiwakoti2015,
    author = {Nirajan Shiwakoti and Yanshan Gong and Xiaomeng Shi and Zhirui Ye},
    title = {Examining influence of merging architectural features on pedestrian crowd movement},
    journal = {Safety Science},
    volume = {75},
    pages = {15-22},
    year = {2015},
    issn = {0925-7535},
    doi = {https://doi.org/10.1016/j.ssci.2015.01.009},
    url = {https://www.sciencedirect.com/science/article/pii/S0925753515000107},
}

@article{Shahhoseini2018,
    author = {Zahra Shahhoseini and Majid Sarvi and Meead Saberi},
    title = {Pedestrian crowd dynamics in merging sections: Revisiting the “faster-is-slower” phenomenon},
    journal = {Physica A: Statistical Mechanics and its Applications},
    volume = {491},
    pages = {101-111},
    year = {2018},
    issn = {0378-4371},
    doi = {https://doi.org/10.1016/j.physa.2017.09.003},
    url = {https://www.sciencedirect.com/science/article/pii/S0378437117308956},
}

@article{zhang2012,
    author={Jun Zhang and Wolfram Klingsch and Tobias Rupprecht and Andreas Schadschneider and Armin Seyfried},
    title={Empirical study of turning and merging of pedestrian streams in T-junction},
    year={2012},
    eprint={1112.5299},
    archivePrefix={arXiv},
    primaryClass={physics.soc-ph},
    doi = {https://doi.org/10.48550/arXiv.1112.52993},
    url = {https://arxiv.org/abs/1112.5299}, 
}

@article{Kayvan2015,
    author = {Aghabayk, Kayvan and Sarvi, Majid and Ejtemai, Omid and Sobhani, Amir},
    year = {2015},
    month = {09},
    pages = {76-83},
    title = {Impacts of Different Angles and Speeds on Behavior of Pedestrian Crowd Merging},
    volume = {2490},
    journal = {Transportation Research Record: Journal of the Transportation Research Board},
    doi = {10.3141/2490-09}
}

@article{Lian2017,
    author = {Liping Lian and Xu Mai and Weiguo Song and Yuen Kwok Kit Richard and Ye Rui and Sha Jin},
    title = {Pedestrian merging behavior analysis: An experimental study},
    journal = {Fire Safety Journal},
    volume = {91},
    pages = {918-925},
    year = {2017},
    issn = {0379-7112},
    doi = {https://doi.org/10.1016/j.firesaf.2017.04.015},
    url = {https://www.sciencedirect.com/science/article/pii/S0379711217301066},
}

@article{Yu2018,
    author = {Hanchen Yu and Nan Jiang and Hongyun Yang and Jixin Shi and Zhenyu Han and Eric Wai Ming Lee and Lizhong Yang},
    title = {Empirical analysis of pedestrian merging process with different merging angles and merging layouts},
    journal = {Physica A: Statistical Mechanics and its Applications},
    volume = {656},
    pages = {130218},
    year = {2024},
    issn = {0378-4371},
    doi = {https://doi.org/10.1016/j.physa.2024.130218},
    url = {https://www.sciencedirect.com/science/article/pii/S0378437124007271},
}

@book{fruin1971,
  title={Pedestrian Planning and Design},
  author={Fruin, J.J.},
  lccn={70159312},
  url={https://books.google.co.jp/books?id=AydSAAAAMAAJ},
  year={1971},
  publisher={Metropolitan Association of Urban Designers and Environmental Planners}
}

@article{Feliciani2018,
    author = {Claudio Feliciani and Katsuhiro Nishinari},
    title = {Measurement of congestion and intrinsic risk in pedestrian crowds},
    journal = {Transportation Research Part C: Emerging Technologies},
    volume = {91},
    pages = {124-155},
    year = {2018},
    issn = {0968-090X},
    doi = {https://doi.org/10.1016/j.trc.2018.03.027},
    url = {https://www.sciencedirect.com/science/article/pii/S0968090X18304133},
}

@article{Zanlungo2023,
    author = {Francesco Zanlungo and Claudio Feliciani and Zeynep Yücel and Xiaolu Jia and Katsuhiro Nishinari and Takayuki Kanda},
    title = {A pure number to assess “congestion” in pedestrian crowds},
    journal = {Transportation Research Part C: Emerging Technologies},
    volume = {148},
    pages = {104041},
    year = {2023},
    issn = {0968-090X},
    doi = {https://doi.org/10.1016/j.trc.2023.104041},
    url = {https://www.sciencedirect.com/science/article/pii/S0968090X2300030X},
}

@article{Guo2020,
    author = {Hui Guo and Xinyao Guo and Wei Lv and Yinghua Song},
    title = {Investigation of crowd’s aggregation measurement based on an entropy model},
    journal = {Safety Science},
    volume = {127},
    pages = {104714},
    year = {2020},
    issn = {0925-7535},
    doi = {https://doi.org/10.1016/j.ssci.2020.104714},
    url = {https://www.sciencedirect.com/science/article/pii/S0925753520301119},
}

@article{Huang2015,
   author = {Lida Huang and Tao Chen and Yan Wang and Hongyong Yuan},
    issn = {0378-4371},
   journal = {Physica A: Statistical Mechanics and its Applications},
   month = {12},
   pages = {200-209},
   publisher = {North-Holland},
   title = {Congestion detection of pedestrians using the velocity entropy: A case study of Love Parade 2010 disaster},
    volume = {440},
    doi = {10.1016/J.PHYSA.2015.08.013},
    year = {2015}
}

@article{Xie2022,
   author = {Wei Xie and Eric Wai Ming Lee and Yiu Yin Lee},
    issn = {0926-5805},
   journal = {Automation in Construction},
   month = {2},
   pages = {104100},
   publisher = {Elsevier},
   title = {Simulation of spontaneous leader–follower behaviour in crowd evacuation},
   volume = {134},
   doi = {10.1016/J.AUTCON.2021.104100},
   year = {2022}
}

@article{Zeng2021,
   author = {Yiping Zeng and Rui Ye and Weiguo Song and Shengfeng Luo and Fanyu Meng and Giuseppe Vizzari},
   issn = {0378-4371},
   journal = {Physica A: Statistical Mechanics and its Applications},
   month = {3},
   pages = {125655},
   publisher = {North-Holland},
   title = {Entropy analysis of the laminar movement in bidirectional pedestrian flow},
   volume = {566},
   doi = {10.1016/J.PHYSA.2020.125655},
   year = {2021}
}

@article{Wang2018,
    author = {J. Wang},
    title = {Research on the Characteristics of Pedestrian Motion and Risk Evaluation Method of the Dense Crowd},
    journal = {STsinghua University PhD Thesis},
    year = {2018},
}

@article{Basalamah2023,
    author = {Saleh Basalamah and Sultan Daud Khan and Emad Felemban and Atif Naseer and Faizan Ur Rehman},
    title = {Deep learning framework for congestion detection at public places via learning from synthetic  data},
    journal = {Journal of King Saud University - Computer and Information Sciences},
    volume = {35},
    number = {1},
    pages = {102-114},
    year = {2023},
    issn = {1319-1578},
    doi = {https://doi.org/10.1016/j.jksuci.2022.11.005},
    url = {https://www.sciencedirect.com/science/article/pii/S1319157822004037},
}

@article{Steffen2010,
    author = {B. Steffen and A. Seyfried},
    title = {Methods for measuring pedestrian density, flow, speed and direction with minimal scatter},
    journal = {Physica A: Statistical Mechanics and its Applications},
    volume = {389},
    number = {9},
    pages = {1902-1910},
    year = {2010},
    issn = {0378-4371},
    doi = {https://doi.org/10.1016/j.physa.2009.12.015},
    url = {https://www.sciencedirect.com/science/article/pii/S0378437109010115},
}

@article{Jia2022,
    author = {Xiaolu Jia and Claudio Feliciani and Hisashi Murakami and Akihito Nagahama and Daichi Yanagisawa and Katsuhiro Nishinari},
    title = {Revisiting the level-of-service framework for pedestrian comfortability: Velocity depicts more accurate perceived congestion than local density},
    journal = {Transportation Research Part F: Traffic Psychology and Behaviour},
    volume = {87},
    pages = {403-425},
    year = {2022},
    issn = {1369-8478},
    doi = {https://doi.org/10.1016/j.trf.2022.04.007},
    url = {https://www.sciencedirect.com/science/article/pii/S1369847822000730},
}

@article{Jia2019,
   author = {Xiaolu Jia and Claudio Feliciani and Daichi Yanagisawa and Katsuhiro Nishinari},
   issn = {0378-4371},
   journal = {Physica A: Statistical Mechanics and its Applications},
   month = {10},
   pages = {121735},
   publisher = {North-Holland},
   title = {Experimental study on the evading behavior of individual pedestrians when confronted with an obstacle in a corridor},
   volume = {531},
   doi = {10.1016/J.PHYSA.2019.121735},
   year = {2019}
}

@article{Courtine2003,
    author = {Courtine, Grégoire and Schieppati, Marco},
    title = {Human walking along a curved path. I. Body trajectory, segment orientation and the effect of vision},
    journal = {European Journal of Neuroscience},
    volume = {18},
    number = {1},
    pages = {177-190},
    doi = {https://doi.org/10.1046/j.1460-9568.2003.02736.x},
    url = {https://onlinelibrary.wiley.com/doi/abs/10.1046/j.1460-9568.2003.02736.x},
    year = {2003}
}

@article{Ye2019,
    title = {Experimental study of pedestrian flow through right-angled corridor: uni- and bidirectional scenarios},
    author = {Rui Ye and Mohcine Chraibi and Chi Liu and Liping Lian and Yiping Zeng and Jun Zhang and Weiguo Song},
    journal = {Journal of Statistical Mechanics Theory and Experiment},
    year = {2019},
    publisher = {IOP Publishing and SISSA},
    number = {4},
    pages = {043401},
    doi = {10.1088/1742-5468/ab0c13},
}

@article{Dias2014,
    author = {Charitha Dias and Omid Ejtemai and Majid Sarvi and Nirajan Shiwakoti},
    title ={Pedestrian Walking Characteristics through Angled Corridors: An Experimental Study},
    journal = {Transportation Research Record},
    volume = {2421},
    number = {1},
    pages = {41-50},
    year = {2014},
    doi = {10.3141/2421-05},
}

@article{Hannun2022,
    doi = {10.1371/journal.pone.0264635},
    author = {Hannun, Jamal AND Dias, Charitha AND Taha, Alaa Hasan AND Almutairi, Abdulaziz AND Alhajyaseen, Wael AND Sarvi, Majid AND Al-Bosta, Salim},
    journal = {PLOS ONE},
    publisher = {Public Library of Science},
    title = {Pedestrian flow characteristics through different angled bends: Exploring the spatial variation of velocity},
    year = {2022},
    month = {03},
    volume = {17},
    url = {https://doi.org/10.1371/journal.pone.0264635},
    pages = {1-21},
}

@article{Duives2015,
    title = {Quantification of the level of crowdedness for pedestrian movements},
    author = {Dorine C. Duives and Winnie Daamen and Serge P. Hoogendoorn},
    journal = {Physica A: Statistical Mechanics and its Applications},
    volume = {427},
    pages = {162-180},
    year = {2015},
    issn = {0378-4371},
    doi = {https://doi.org/10.1016/j.physa.2014.11.054},
}

@article{Wong2010,
    author = {S. C. Wong  and W. L. Leung  and S. H. Chan  and William H. K. Lam  and Nelson H. C. Yung  and C. Y. Liu  and Peng Zhang },
    title = {Bidirectional Pedestrian Stream Model with Oblique Intersecting Angle},
    journal = {Journal of Transportation Engineering},
    volume = {136},
    number = {3},
    pages = {234-242},
    year = {2010},
    doi = {10.1061/(ASCE)TE.1943-5436.0000086},
}

@article{Sharifi2020,
    title = {Exploring heterogeneous pedestrian stream characteristics at walking facilities with different angle intersections},
    author = {Mohammad Sadra Sharifi and Ziqi Song and Hossein Nasr Esfahani and Keith Christensen},
    journal = {Physica A: Statistical Mechanics and its Applications},
    volume = {540},
    pages = {123112},
    year = {2020},
    issn = {0378-4371},
doi = {https://doi.org/10.1016/j.physa.2019.123112},
}

@article{Vleuten2024,
  title = {Stochastic fluctuations of diluted pedestrian dynamics along curved paths},
  author = {van der Vleuten, Geert G. M. and Toschi, Federico and Schilders, Wil and Corbetta, Alessandro},
  journal = {Phys. Rev. E},
  volume = {109},
  issue = {1},
  pages = {014605},
  numpages = {14},
  year = {2024},
  month = {Jan},
  publisher = {American Physical Society},
  doi = {10.1103/PhysRevE.109.014605},
  url = {https://link.aps.org/doi/10.1103/PhysRevE.109.014605}
}

@article{Yu2026,
    author = {Hanchen Yu and Eric Wai Ming Lee and Nan Jiang and Jixin Shi and Weiheng Xie and Hongyun Yang and Lizhong Yang},
    title = {Analysis of pairwise pedestrian collision avoidance dynamics in T-junctions with different merging setups},
    journal = {Chaos, Solitons \& Fractals},
    volume = {203},
    pages = {117674},
    year = {2026},
    issn = {0960-0779},
    doi = {https://doi.org/10.1016/j.chaos.2025.117674},
    url = {https://www.sciencedirect.com/science/article/pii/S096007792501687X},
}

@article{petrack,
title = {Collecting pedestrian trajectories},
journal = {Neurocomputing},
author = {Maik Boltes and Armin Seyfried},
volume = {100},
pages = {127 - 133},
year = {2013},
note = {Special issue: Behaviours in video},
issn = {0925-2312},
}

@misc{Feliciani2016,
author = {Claudio Feliciani and Kenichiro Shimura and Katsuhiro Nishinari},
title = {Turning and merging in curved corridors at various densities and angles},
howpublished = {http://ped.fz-juelich.de/extda/feliciani2016a},
doi = {https://doi.org/10.34735/ped.2016.3},
note = {{Dataset currently unlisted in the main page, will be listed upon acceptance}}
}

\clearpage

\begin{titlepage}
\thispagestyle{empty}
\raggedright 

{\Large\bfseries Conflict Avoidance in Pedestrian Merging in Controlled Experiments by Variance Indicator\par}

\vspace{1em}

{\normalsize
Jiawei Zhang$^{1,*}$,
Xiaolu Jia$^{2}$,
Sakurako Tanida$^{3}$,
Claudio Feliciani$^{3}$,
Daichi Yanagisawa$^{3,1,5}$,
Katsuhiro Nishinari$^{3,1,4,5}$
\par}

\vspace{0.8em}

{\small
$^{1}$ Department of Advanced Interdisciplinary Studies, School of Engineering, The University of Tokyo,
4-6-1 Komaba, Meguro-ku, Tokyo 153-8904, Japan\par
$^{2}$ Beijing Key Laboratory of Traffic Engineering, Beijing University of Technology,
No. 100 Pingleyuan, Chaoyang-district, Beijing 100124, China\par
$^{3}$ Department of Aeronautics and Astronautics, School of Engineering, The University of Tokyo,
7-3-1 Hongo, Bunkyo-ku, Tokyo 113-8656, Japan\par
$^{4}$ Research Center for Advanced Science and Technology, The University of Tokyo,
4-6-1 Komaba, Meguro-ku, Tokyo 153-8904, Japan\par
$^{5}$ Mobility Innovation Collaborative Research Organization, The University of Tokyo, Tokyo, Japan\par
}

\vspace{0.8em}

{\small
$^{*}$ Corresponding author: \texttt{zhang-jiawei@g.ecc.u-tokyo.ac.jp}\par
}

\vspace{1.2em}

\vspace{1.0em}

\noindent\textbf{Acknowledgement}\par
\vspace{0.3em}
This work was supported by JST SPRING (JPMJSP2108), JST-Mirai Program (JPMJMI20D1), Japan,
the National Natural Science Foundation of China (No.~52402376),
JSPS KAKENHI (JP23K21019, JP23K13521, JP25K17790, JP23K20947),
the Odakyu Foundation Research Program, and the Obayashi Foundation Research Program (ST).
In completing this manuscript, the authors used OpenAI's ChatGPT to refine the language, check for writing errors,
and ensure grammatical accuracy.

\end{titlepage}

\clearpage
\begin{titlepage}
\thispagestyle{empty}
\raggedright

\noindent{\Large\bfseries CRediT authorship contribution statement\par}
\vspace{0.8em}

Jiawei Zhang: Conceptualization, Formal analysis, Data curation, Funding acquisition, Methodology, Software,
Visualization, Writing -- original draft.\par
\vspace{0.4em}
Xiaolu Jia: Funding acquisition, Supervision, Writing -- review.\par
\vspace{0.4em}
Sakurako Tanida: Funding acquisition, Supervision, Writing -- review.\par
\vspace{0.4em}
Claudio Feliciani: Data curation, Funding acquisition, Investigation, Methodology, Supervision, Writing -- review.\par
\vspace{0.4em}
Daichi Yanagisawa: Funding acquisition, Supervision, Writing -- review and editing.\par
\vspace{0.4em}
Katsuhiro Nishinari: Conceptualization, Funding acquisition, Supervision.\par

\end{titlepage}

\clearpage

\end{document}